\documentclass[aps,prl,reprint,superscriptaddress,footinbib]{revtex4-1}

\usepackage[T1]{fontenc}       % Use modern font encodings
\usepackage{amsmath,amssymb}
\usepackage{graphicx}
\usepackage{braket}            % Use for the average (<...>) symbol.
\usepackage{upgreek}           % Use for upright Greek letters
\usepackage{color}
\usepackage{booktabs} % Use for nice tables.
\usepackage[utf8x]{inputenc}
\renewcommand{\vec}[1]{\mathbf{#1}} % Use bold for vectors.

\begin{document}

\title{Twist-bend coupling and the torsional response of double-stranded
DNA}

\author{Stefanos K.\ Nomidis}
\affiliation{KU Leuven, Institute for Theoretical Physics, 
Celestijnenlaan 200D, 3001 Leuven, Belgium}
\affiliation{Flemish Institute for Technological Research (VITO),
Boeretang 200, B-2400 Mol, Belgium}

\author{Franziska Kriegel}
\affiliation{Department of Physics, Nanosystems Initiative Munich,
and Center for NanoScience, LMU Munich, 80799~Munich, Germany}

\author{Willem Vanderlinden}
\affiliation{Department of Physics, Nanosystems Initiative Munich, and Center 
for NanoScience, LMU Munich, Amalienstrasse 54, 80799 Munich, Germany}
\affiliation{KU Leuven, Division of Molecular Imaging and Photonics, 
Celestijnenlaan 200F, 3001 Leuven, Belgium}

\author{Jan Lipfert}
\affiliation{Department of Physics, Nanosystems Initiative Munich, and Center 
for NanoScience, LMU Munich, Amalienstrasse 54, 80799 Munich, Germany}

\author{Enrico Carlon}
\affiliation{KU Leuven, Institute for Theoretical Physics, 
Celestijnenlaan 200D, 3001 Leuven, Belgium}

\begin{abstract}
Recent magnetic tweezers experiments have reported systematic deviations
of the twist response of double-stranded DNA from the predictions of the
twistable worm-like chain model. Here we show, by means of analytical
results and computer simulations, that these discrepancies can be resolved
if a coupling between twist and bend is introduced. We obtain an estimate
of $40\pm10$~nm for the twist-bend coupling constant. Our simulations are
in good agreement with high-resolution, magnetic-tweezers torque data.
Although the existence of twist-bend coupling was predicted long ago
(Marko and Siggia, Macromolecules {\bf 27}, 981 (1994)), its effects on
the mechanical properties of DNA have been so far largely unexplored. We
expect that this coupling plays an important role in several aspects
of DNA statics and dynamics.
\end{abstract}

\date{\today}

\maketitle

\begin{figure}[b]
\centering\includegraphics[width=0.42\textwidth]{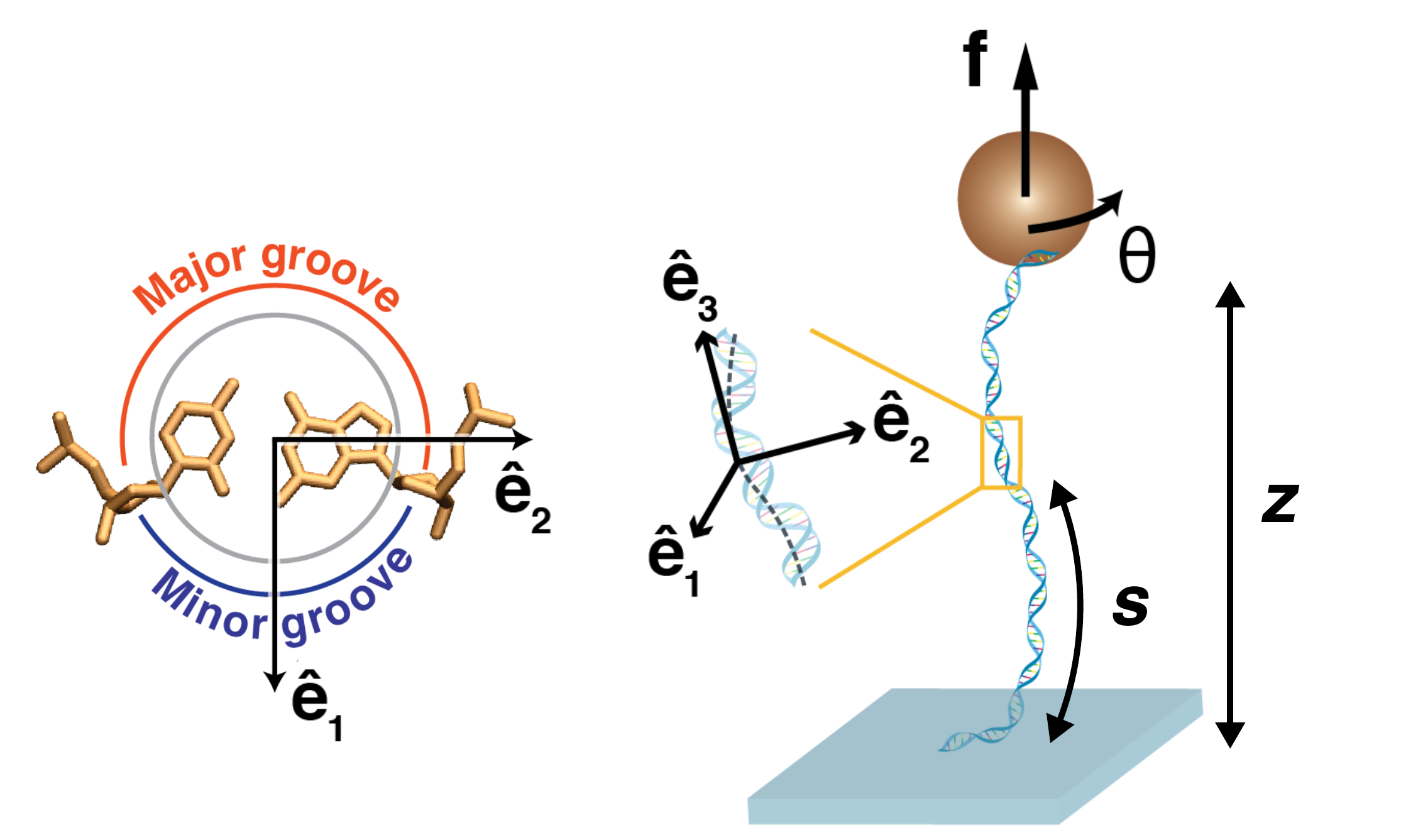}
\caption{Schematic representation of a typical MT experiment. Magnetic fields 
are used to apply forces and torques (inducing a rotation angle $\theta$) to a 
paramagnetic bead. A dsDNA molecule is attached at one side to the bead and at 
the other to a flow-cell surface, separated by a distance $z$ measuring the 
extension of the molecule. Continuum elastic models describe the double-helix 
conformation using an orthonormal frame $\{ \vec{\widehat{e}}_1, 
\vec{\widehat{e}}_2, \vec{\widehat{e}}_3 \}$ at each point along the molecule, 
labeled by a coordinate $s$. $\vec{\widehat{e}}_3$ is tangent to the helical 
axis, while $\vec{\widehat{e}}_1$ points from the center of the helix towards 
the middle of the minor groove, and $\vec{\widehat{e}}_2 = \vec{\widehat{e}}_3 
\times \vec{\widehat{e}}_1$.}
\label{Fig:MT}
\end{figure}

\paragraph{Introduction} The mechanical properties of double-stranded
DNA (dsDNA) are critical for both its structure and function within
the cell. The stretching of dsDNA under applied forces has been
measured by single molecule techniques \cite{smit92,bust03} and is
accurately reproduced by a simple polymer model, containing the bending
stiffness as the only parameter~\cite{bust94}. Elastic polymer models
were also successfully employed to study the torsional properties of
dsDNA~\cite{mark94b} and compared to single-molecule experiments, such
as magnetic tweezers (MT) \cite{stri96} (Fig.~\ref{Fig:MT}, right). The
currently accepted elastic model for dsDNA is the twistable worm-like
chain (TWLC)~\cite{nels08}. Although the TWLC correctly describes the
overall response of dsDNA to applied forces and torques, it fails to
quantitatively explain the force-dependence of the effective torsional
stiffness~\cite{lipf10,lipf11}.  Here, we show that an alternative elastic
model proposed by Marko and Siggia (MS)~\cite{mark94}, quantitatively
describes the force-dependence of the effective torsional stiffness, by
taking into account a direct coupling between twist and bend deformations.
Furthermore, we demonstrate that the MS model explains an unresolved
discrepancy in the measured \emph{intrinsic} torsional stiffness,
obtained from different techniques. Finally,  we show that the MS model
provides a better description of the pre-buckling torque response of
dsDNA, determined in high-resolution magnetic torque tweezers (MTT)
experiments, than the TWLC.

%%%%%%%%%%%%%%%%%%%%%%%%%%%%%%%%%%%%%%%%%%%%%%%%%%%%%%%%%%%%%%%%
%%%%%%%%%%%%%%%%%%%%%%%%%%%%%%%%%%%%%%%%%%%%%%%%%%%%%%%%%%%%%%%%

\paragraph{TWLC and MS models} 
Both the TWLC and MS models describe dsDNA as a continuous, twistable
curve by associating an orthonormal frame $\{ \vec{\widehat{e}}_1,
\vec{\widehat{e}}_2, \vec{\widehat{e}}_3 \}$ with each base pair
(Fig.~\ref{Fig:MT})~\cite{mark94}. We choose $\vec{\widehat{e}}_3$ tangent
to the helical axis and $\vec{\widehat{e}}_1$ and $\vec{\widehat{e}}_2$
oriented as in Fig.~\ref{Fig:MT}. In the continuum limit these vectors
are functions of the arc-length variable $s$.  For the stretching forces
considered here ($f < 10$~pN) dsDNA is inextensible, hence $0 \leq s
\leq L$, with $L$ the contour length. A local dsDNA conformation is
given by a vector $\vec\Omega (s)$ which describes the infinitesimal
rotation connecting $\{ \vec{\widehat{e}}_1 (s), \vec{\widehat{e}}_2
(s), \vec{\widehat{e}}_3 (s)\}$ to $\{ \vec{\widehat{e}}_1 (s+ds),
\vec{\widehat{e}}_2 (s+ds), \vec{\widehat{e}}_3 (s+ds)\}$.  The direction
of $\vec\Omega (s)$ identifies the rotation axis, and $|\vec\Omega (s)|ds$
the infinitesimal rotation angle. In particular, if $\vec\Omega (s)$ is
parallel to $\vec{\widehat{e}}_3 (s)$, one generates a local rotation
along the tangent vector, i.e.\ a twist deformation. Conversely, an
$\vec\Omega (s)$ along $\vec{\widehat{e}}_1 (s)$ or $\vec{\widehat{e}}_2
(s)$ corresponds to a bending deformation. Expressing the local rotation
vector as $\vec\Omega (s) = \sum_{i=1}^3 \Omega_i(s) \vec{\widehat{e}}_i
(s)$, one identifies the twist mode with $\Omega_3$ and the two bending
modes with $\Omega_1$ and $\Omega_2$.

Marko and Siggia~\cite{mark94} showed that the molecular symmetry of
dsDNA imposes \emph{only} the invariance of the energy to the interchange
$\Omega_1 \to -\Omega_1$. This leads, to lowest order in $\Omega_i$, to the
following energy functional
\begin{equation}
\beta E_\text{MS} = \frac 1 2 \int_0^L ds 
\left(
A_1 \Omega_1^2 + A_2 \Omega_2^2 + C   \Omega_3^2 +
2G \Omega_2 \Omega_3 + \ldots
\right),
\label{eq:model_MS}
\end{equation}
where $\beta \equiv 1/k_BT$, and the dots denote higher-order terms. The MS 
model is characterized by two bending stiffnesses $A_1$ and $A_2$, a torsional 
stiffness $C$ and a twist-bend coupling constant $G$, which 
have dimension of length. Note that twist-bend coupling $\Omega_2 \Omega_3$ is 
the only quadratic cross-term allowed by the $\Omega_1 \to -\Omega_1$ symmetry 
\cite{mark94}. 

The TWLC is a limiting case of the MS model obtained by setting
$A_1 = A_2 \equiv A$ and $G=0$
\begin{equation}
\beta E_\text{TWLC} = \frac 1 2 \int_0^L ds 
\left[ A \left(\Omega_1^2 + \Omega_2^2 \right)+ C \Omega_3^2 
+\ldots \right].
\label{eq:model_TWLC}
\end{equation}
We note that the asymmetric bending stiffness and the twist-bend
coupling, which are intrinsic to the MS model and neglected in the
TWLC, are naturally suggested by the structure of the DNA helix, with
its pronounced minor and major groove (Fig.~\ref{Fig:MT}, left). In
the following, we analyze the consequences of taking these additional
terms into account using analytical calculations and extensive computers
simulations. Although the existence of twist-bend coupling
was predicted long ago, its effects on the mechanical properties of DNA
have been so far largely unexplored. Two studies~\cite{lank00, moha05}
in which the MS model was invoked are discussed below.

\paragraph{Renormalized bending and torsional stiffnesses}
In order to parametrize the MS model, we calculated the
renormalized bending and torsional stiffness, 
$\kappa_b$ and $\kappa_t$ respectively, from the 
equilibrium fluctuations of a free chain (see the 
Supplemental Material~\cite{suppl} for derivation):
\begin{eqnarray}
\kappa_\text{b}&=& A 
\,
\frac{1-\dfrac{\varepsilon^2}{A^2}-
\dfrac{G^2}{AC}\left(1+\dfrac{\varepsilon}{A}\right)}
 {1-\dfrac{G^2}{2AC}},
 \label{eq:lb} \\
\kappa_\text{t}&=& C 
\,
\frac{1-\dfrac{\varepsilon}{A}- \dfrac{G^2}{AC}}
{1-\dfrac{\varepsilon}{A}},
\label{eq:lt}
\end{eqnarray}
where $A \equiv(A_1+A_2)/2$ is the mean bending stiffness and
$\varepsilon \equiv(A_1-A_2)/2$ the bending anisotropy. By setting
$\varepsilon=G=0$ one obtains the TWLC values $\kappa_\text{b} = A$
and $\kappa_\text{t} = C$.  Eqs.~\eqref{eq:lb} and \eqref{eq:lt} show
that in the MS model $\kappa_\text{b} < A$ and $\kappa_\text{t} < C$,
hence the presence of a direct twist-bend coupling softens the chain,
rendering the bending and twisting fluctations larger than expected from
the intrinsic parameters $A$ and $C$, respectively.  The details of the
parametrization are discussed below. Note, finally, that $\kappa_\text{b}$
and $2\kappa_\text{t}$ are also the bending and twisting persistence
lengths, that characterize the decay of the respective correlations
along the molecule.

\paragraph{Effective torsional stiffness} The parameter $C$ in 
Eqs.~\eqref{eq:model_MS} and \eqref{eq:model_TWLC} is the \emph{intrinsic} 
torsional stiffness and quantifies the energetic cost of \emph{local} 
pure twist deformations ($\Omega_1=\Omega_2=0$, 
$\Omega_3 \neq 0$). The \emph{effective} torsional stiffness $C_\text{eff}$, 
in contrast, expresses the cost of a \emph{global} twist deformation, and 
decreases with decreasing force. The force dependence of $C_\text{eff}$ can be 
understood as follows: In absence of thermal fluctuations a weakly twisted dsDNA 
is straight, and the twist response is governed by the intrinsic torsional 
stiffness $C$. In the presence of thermal fluctuations, however, twist can be 
absorbed by bending~\cite{moro97,bouc98}, leading to an effective torsional 
stiffness $C_\text{eff}<C$. High stretching forces suppress bending 
fluctuations, therefore yielding $z \approx L$ and $C_\text{eff} \approx C$, 
while at low forces fluctuations are high, hence $z < L$ and $C_\text{eff} < C$. 
Moroz and Nelson calculated the force dependence of $C_\text{eff}$ for the TWLC 
in the high-force limit~\cite{moro97}
\begin{equation}
C_\text{eff} =  C \,
 \left(1-\frac{C}{4A}\sqrt{\frac{k_BT}{Af}} + \ldots \right),
\label{eq:ceff}
\end{equation}
where the dots indicate higher-order corrections in $1/\sqrt{f}$.

\begin{figure}[t]
\centering\includegraphics[width=\linewidth]{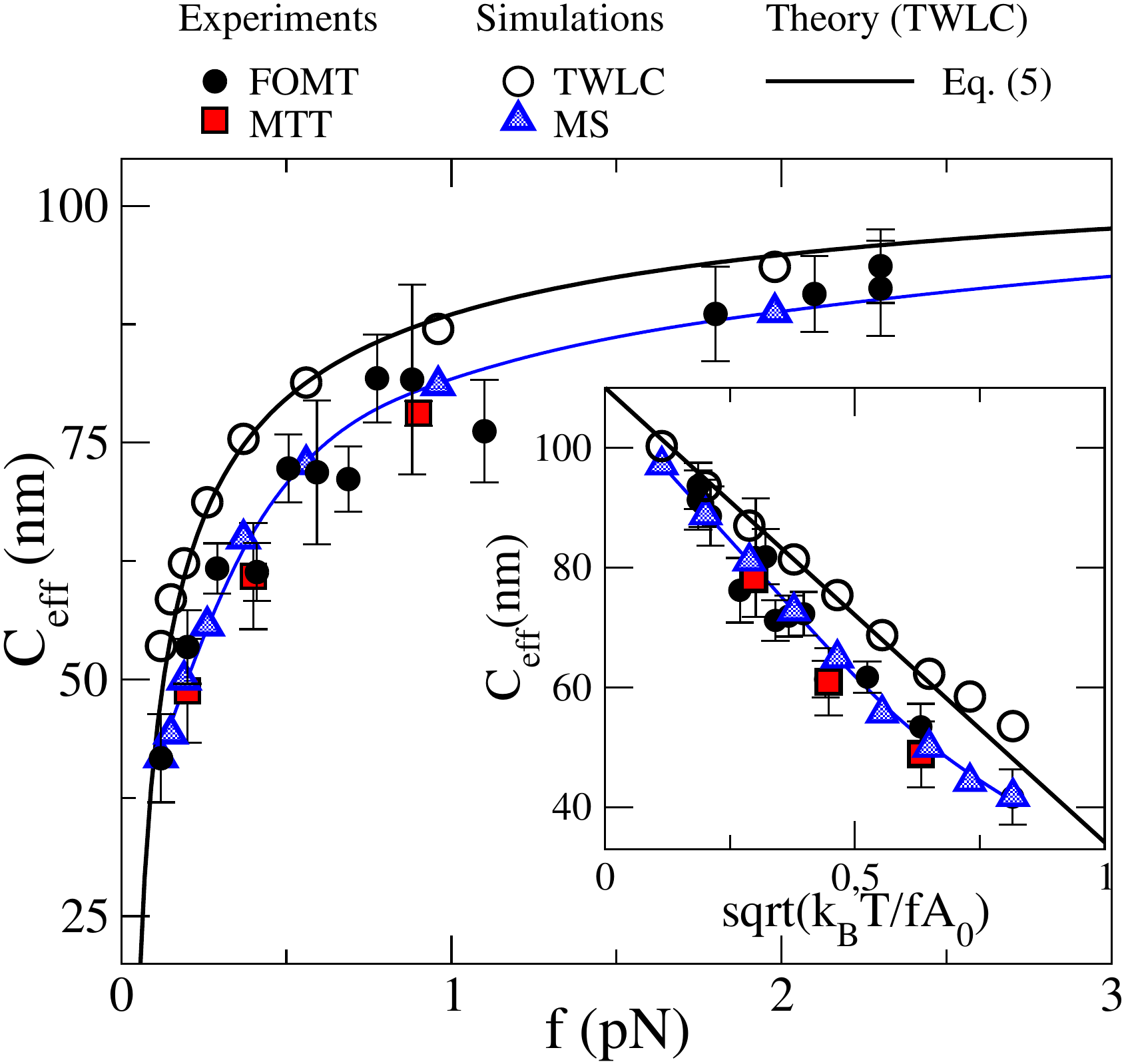} 
\caption{Force-dependence of the effective torsional stiffness from MTT
(present work) and FOMT~\cite{lipf11} measurements, from simulations
of the TWLC and the MS models (where error bars are smaller than the
symbols) and from the analytical TWLC approximation (Eq.~\eqref{eq:ceff}).
Parameters are  $A = 43$~nm and $C = 110$~nm for the TWLC and $A
=56$~nm, $C = 110$~nm, $\varepsilon=10$~nm and $G=40$~nm for the MS model.
The inset shows $C_\text{eff}$ as a function of the rescaled variable
$\sqrt{k_BT/A_0f}$ (with $A_0=50$~nm); in these units Eq.~\eqref{eq:ceff}
becomes a straight line. The experimental data are not well described
by the TWLC, but agree quantitatively with the MS model (reduced
$\chi^2_\text{TWLC} = 6.1$ and $\chi^2_\text{MS} = 0.74$, respectively). The
blue line is an interpolation of the MS simulations points.}
\label{Fig:Ceff}
\end{figure}

$C_\text{eff}$ has been experimentally measured with two independent
single-molecule approaches. In magnetic torque tweezers (MTT) $C_\text{eff}$ is
obtained from the torque response $\tau$ upon over- and underwinding
dsDNA by a small angle $\theta$  (Fig.~\ref{Fig:MT}, 
right)~\cite{lipf10, lipf14, kaue11,ober12}
\begin{equation}
\tau \approx \frac{k_B T C_\text{eff}}{L} \theta.
\label{eq:mtt}
\end{equation}
Freely orbiting magnetic tweezers (FOMT)~\cite{lipf11} and the 
rotor bead assay~\cite{brya03, ober12}, in contrast, measure fluctuations of 
$\theta$ of a freely rotating dsDNA tether, and $C_\text{eff}$ is obtained from 
\begin{equation}
C_\text{eff} = \frac{L}{\sigma_\theta^2},
\label{eq:fomt}
\end{equation}
where $\sigma_\theta^2$ is the variance of $\theta$. MTT and FOMT
yield consistent values of $C_\text{eff}$, which, however, deviate
from the TWLC prediction of Eq.~\eqref{eq:ceff} \cite{lipf10,lipf11}
(Fig.~\ref{Fig:Ceff}).

%%%%%%%%%%%%%%%%%%%%%%%%%%%%%%%%%%%%%%%%%%%%%%%%%%%%%%%%%%%%%%%%
%%%%%%%%%%%%%%%%%%%%%%%%%%%%%%%%%%%%%%%%%%%%%%%%%%%%%%%%%%%%%%%%

To investigate the force-dependence of $C_\text{eff}$, we performed
computer simulations of the TWLC and the MS models using a coarse-grained
model, similar to Ref.~\cite{brac14}, where the dsDNA is represented
by a series of connected beads, each carrying an orthonormal frame of
reference. Successive beads interact via potential energies obtained from
the discretization of Eqs.~\eqref{eq:model_MS} or \eqref{eq:model_TWLC}
in the MS or TWLC models, respectively \cite{suppl}. The effect of an
implicit solvent was introduced via Langevin forces~\cite{zwan01}.

The TWLC simulations with $A=43$~nm and $C=110$~nm (Fig.~\ref{Fig:Ceff},
empty circles) are in excellent agreement with the high-force expansion
of Eq.~\eqref{eq:ceff}, (Fig.~\ref{Fig:Ceff}, black solid line). This is
a good test of our simulations and shows that higher-order corrections
to~\eqref{eq:ceff} do not contribute significantly to $C_\text{eff}$.
The value of $A = 43$~nm comes from the measured persistence length in
Ref.~\cite{lipf11} while the value $C = 110$~nm of the intrinsic torsional
stiffness has been obtained from extrapolation of $C_\text{eff}$ at high
forces (see inset of Fig.~\ref{Fig:Ceff} and~\cite{suppl}).

We then turned to the MS model, which was parametrized as follows:
Similarly to the TWLC, the intrinsic torsional stiffness was set at
$C=110$~nm.  Following Ref.~\cite{noro08}, we chose $\varepsilon=10$~nm
for the bending anisotropy (tests for different values of $\varepsilon$
are shown in the Supplemental Material \cite{suppl}). $A$ and $G$
were chosen so that Eq.~\eqref{eq:lb} gives $\kappa_\text{b}=43$~nm,
the measured persistence length of dsDNA.  Therefore only one of the
two could be freely adjusted. The best fit to the data was found for
$G=40$~nm and $A = 56$~nm, and is in quantitative agreement with the
experiments.  Control simulations for other values of $\varepsilon$ and
$\kappa_\text{b}$ gave similar estimates of the twist-bend stiffness
\cite{suppl}. We finally obtained $G = 40\pm10$~nm, where the error
covers the range of values for which simulations fit the MT data within
their experimental errors.

This value is somewhat higher than the estimate $G = 25$~nm~\cite{moha05},
obtained from the analysis of structural correlations of dsDNA wrapped
around a histone core. Elastic couplings in dsDNA were also investigated
in all-atom simulations~\cite{lank00,lank03}. These and more recent
studies~\cite{drvs14} show that the twist-bend coupling is the most
significant among the off-diagonal elastic terms (i.e.\ $\Omega_1
\Omega_2$, $\Omega_1 \Omega_3$ and $\Omega_2\Omega_3$), in agreement
with the symmetry analysis by Marko and Siggia \cite{mark94}.

\paragraph{Intrinsic torsional stiffness} The experimental determination
of the intrinsic torsional stiffness~$C$ has proven to be a challenging
task, with different experimental techniques yielding a wide range of
values as $40-120$~nm~\cite{shib85, fuji90, lipf10, kaue11, brya12,
mosc09, fort08, moro98, bouc98}. The techniques used for this purpose
can be divided into two main categories. The first group contains
single-molecule techniques as MT, in which a stretching force is
applied to DNA.  One can obtain $C$ from high-force extrapolation
of $C_\text{eff}$, which typically yields values in the range
$100-110$~nm~\cite{mosc09, fort08, moro98, lipf10}. In the second group
of techniques no force is applied to the DNA molecule, as in fluorescence
polarization anisotropy~\cite{shib85,fuji90}, the analysis of cyclization
rates~\cite{leve86} or topoisomer distributions~\cite{shor83}. Typical
values from these studies are lie in the range $60-80$~nm~\cite{suppl}.

According to the TWLC all the above techniques should probe the intrinsic
torsional stiffness $C$. In the framework of the MS model, however,
this is not the case; the torsional response at high tension is still
governed by $C$, since in this limit bending fluctuations are suppressed,
i.e.\ $\Omega_1$, $\Omega_2 \to 0$. In absence of applied forces, however
bending fluctuations influence the measured torsional stiffness via the
twist-bend coupling $G$, leading to twist stiffness $\kappa_\text{t}<C$
according to Eq.~\eqref{eq:lt}).  With the parametrization used for
the fit of Fig.~\ref{Fig:Ceff} ($A = 56$~nm, $\varepsilon = 10$~nm, $C =
110$~nm and $G = 40$~nm), Eq.~\eqref{eq:lt} gives $\kappa_\text{t}=75$~nm,
which is consistent with the values obtained from the second family of
techniques (details in Supplemental Material~\cite{suppl}).  We conclude
that, the wide spread in the experimental $C$ values, which appears to
be a discrepancy in the TWLC model, is naturally explained within the
framework of the MS model.

%%%%%%%%%%%%%%%%%%%%%%%%%%%%%%%%%%%%%%%%%%%%%%%%%%%%%%%%%%%%%%%%
\begin{figure}[t]
\centering\includegraphics[width=\linewidth]{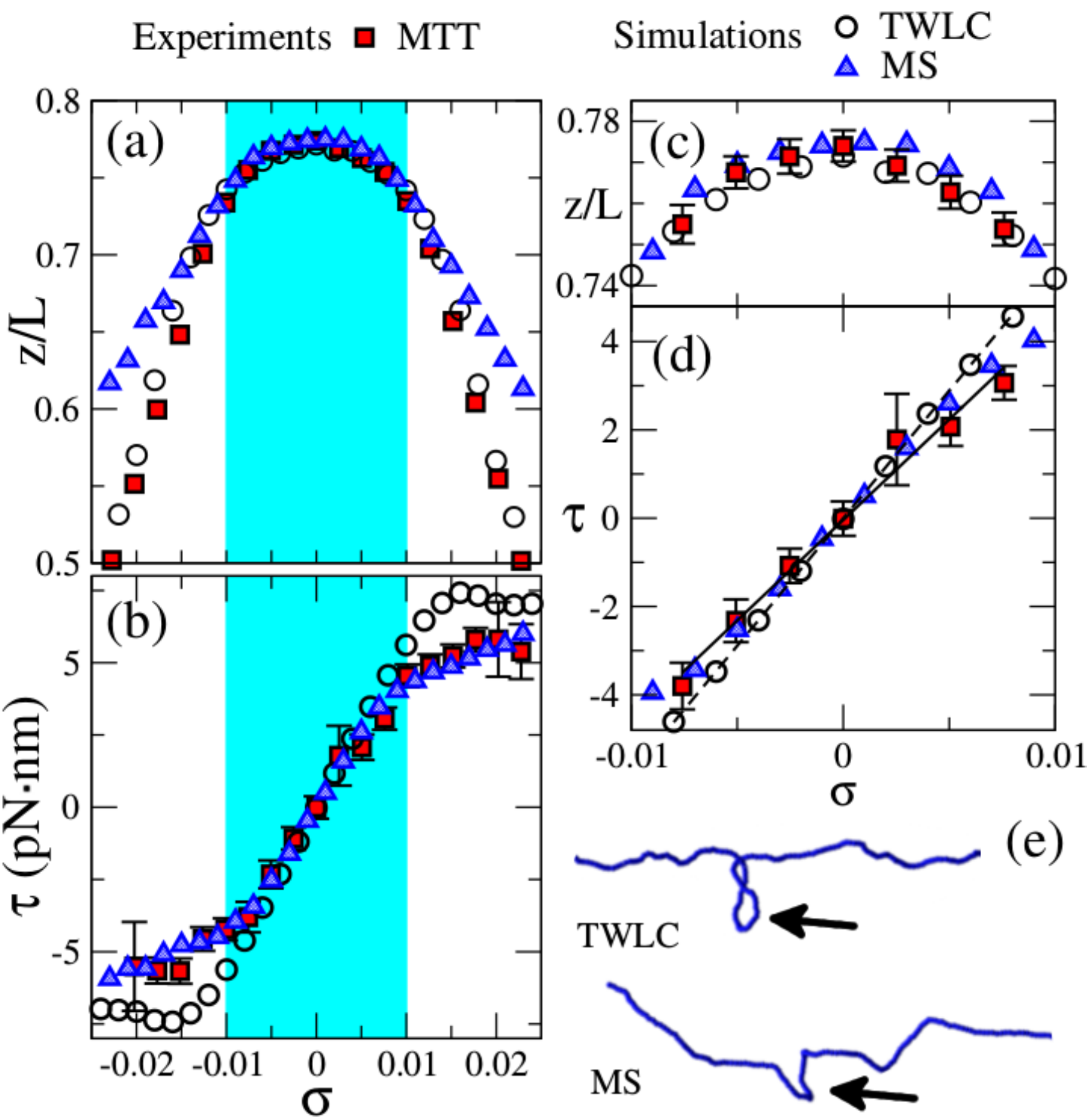}
\caption{(a) Relative extension $z/L$ and (b) torque $\tau$
vs. supercoiling density $\sigma$ at $f=0.4$~pN from MTT experiments
(filled squares) and simulations of the TWLC (empty circles, $A=43$~nm,
$C=110$~nm) and MS (filled triangles, $A=56$~nm, $C=110$~nm, $G=40$~nm
and $\varepsilon=10$~nm) models.  (c) and (d) are zooms of $z/L$ and
$\tau$ in the pre-buckling regime, shaded area in (a) and (b). (e)
Snapshots of simulations of the TWLC and MS models, respectively, at
$\sigma=0.023$. The arrows point to a plectonemic supercoil in the TWLC
and to a solenoidal supercoil in the MS model.}
\label{Fig:torq_ext}
\end{figure}
%%%%%%%%%%%%%%%%%%%%%%%%%%%%%%%%%%%%%%%%%%%%%%%%%%%%%%%%%%%%%%%%

\paragraph{Extension and torque} To further investigate the effect of
twist-bend coupling, we performed high-resolution MTT extension and torque
measurements and compared to simulations of the TWLC and MS models. We
introduce the supercoiling density $\sigma \equiv {\theta}/{(\omega_0
L)}$, where $\omega_0 \approx 1.85$~nm$^{-1}$ is the intrinsic twist
density of dsDNA. Although the discussion so far has been limited to
the regime of small $|\sigma|$, we now extend our analysis to high
values of $|\sigma|$, for which dsDNA undergoes a buckling transition
to interwound structures called plectonemes. We will first discuss
the pre-buckling (low $|\sigma|$) and then the post-buckling (high
$|\sigma|$) regime.

Fig.~\ref{Fig:torq_ext}(a) and (b) show the $\sigma$-dependence of the
relative extension~$z/L$ and torque~$\tau$, respectively, for $f = 0.4$~pN
(similar plots for $f=0.2$ and $0.9$~pN are in the Supplemental Material
\cite{suppl}). The shaded area in Fig.~\ref{Fig:torq_ext} indicates
the pre-buckling regime, in which the simulated values of $z/L$ for the
two models differ by less than $1\%$ (Fig.~\ref{Fig:torq_ext}(c)). In
addition, both models are consistent with the MTT measurements of the
dsDNA extension (reduced $\chi^2_\text{TWLC} = 1.6$ and $\chi^2_\text{MS}
= 0.77$, averaged over all forces).  In contrast, there is a noticeable
difference between the models in the behavior of the torque at small
$|\sigma|$ (Fig.~\ref{Fig:torq_ext}(d)), with the MS model providing a
significantly better prediction of the experimental data than the TWLC
(reduced $\chi^2_\text{TWLC} = 5.9$ and $\chi^2_\text{MS} = 1.3$, again
averaged over all forces).  The quantitative agreement of the MS model
with experimental extension and torque data in the pre-buckling regime
further highlights the importance of including the twist-bend coupling
term in elastic models of dsDNA, and provides an independent test of
our estimate of $G$.

We now focus on the post-buckling regime, where simulations of the
TWLC show that plectonemes form at sufficiently high $|\sigma|$
(Fig.~\ref{Fig:torq_ext}(e)) and that the relative extension data
fit well the experiments (Fig.~\ref{Fig:torq_ext}(a)), in agreement
with previous work \cite{scho12, lepa15, iven16}. However the TWLC
fails to account for the torque data, both in the pre-buckling,
as already discussed, and even more clearly in the post-buckling
regime (Fig.~\ref{Fig:torq_ext}(b)). Our work highlights the need
for direct torque measurements to quantitatively test elastic
models of DNA. Earlier comparisons between analytical predictions
and experiments were limited to extension data~\cite{moro98,bouc98},
since only in the past few years direct torque measurements have become
available~\cite{lipf10,kaue11,brya03,ober12,fort08}.

The post-buckling behavior of the MS is somehow complementary to that of
the TWLC: The torque is in quantitative agreement with the experiments,
but there are deviations in the extension. Fig.~\ref{Fig:torq_ext}(e)
shows simulation snapshots of equilibrium conformations for the TWLC
and MS models at $\sigma=0.023$. Whereas the TWLC forms a plectoneme,
the MS model favors highly-twisted helical configurations, known as
solenoidal supercoils~\cite{mark94}. This different behavior can be
explained as follows: Upon twisting, there is an energetic penalty in
the TWLC, due to $C\Omega_3^2$ in Eq.~\eqref{eq:model_TWLC}.  Beyond a
threshold value of $|\sigma|$, part of the twist is transformed into
a localized bending deformation, giving rise to plectonemic supercoils
(Fig.~\ref{Fig:torq_ext}(e)). On the other hand, in
the MS model~\eqref{eq:model_MS}, the torsional energy can be reduced by
a uniform bending such that the quantity $\braket{\Omega_2 \Omega_3 }$
becomes negative. This gives rise to the characteristic helical structures
(solenoidal supercoils, Fig.~\ref{Fig:torq_ext}(e)).  The absence
of plectonemes in the range of $\sigma$ considered is a shortcoming of
the MS model, as defined by the energy functional~\eqref{eq:model_MS},
and is the reason why $z/L$ decreases with $|\sigma|$ less steeply than
in the experiments.

Thus, we conclude that at high $\sigma$ both models deviate from
experiments, although in different ways.  It should be stressed that
both models are obtained as quadratic expansions in the deformation
parameters $\Omega_i$. It is likely that, close to buckling, higher-order
anharmonic terms in $\Omega_i$ will become relevant. This is particularly
true for the MS model, where the molecular asymmetry of dsDNA allows
six additional third-order terms: $\Omega_1^2 \Omega_2$, $\Omega_1^2
\Omega_3$, $\Omega_2^2 \Omega_3$, $\Omega_2 \Omega_3^2$, $\Omega_2^3$
and $\Omega_3^3$ \cite{mark94}.

\paragraph{Discussion} We investigated the mechanical properties of an elastic 
DNA model with an explicit twist-bend coupling~\cite{mark94}. Our analysis 
focused on the regime where the supercoiling density is small, i.e.\ twist and 
bending deformations are weak. We showed that the model resolves two issues that 
the standard TWLC fails to explain: i) the force-dependence of the effective 
torsional stiffness, also discussed in the recent 
literature~\cite{lipf10,lipf11,schu15,lepa15} and ii) discrepancies in the 
reported estimates of the intrinsic torsional 
stiffness~\cite{moro98,shib85,fuji90}. Moreover, it provides a 
superior fit to high-resolution torque data, compared to the TWLC.

An alternative model, that invokes a cooperative structural 
transition in the DNA helix, was recently proposed by Schurr~\cite{schu15}. 
While this model also explains the deviations between the $C_\text{eff}$ data 
and the TWLC predictions, we note that our current model naturally follows from 
the molecular symmetry of the DNA helix \cite{mark94} and independently explains
several different features of the torsional response of DNA.

Although we showed that the MS model is a more accurate mechanical model of 
dsDNA than the standard TWLC in the pre-buckling regime, there is still an open 
issue at high $|\sigma|$, where neither of the two models fully agrees with the 
experiments. The MS model, as defined by the energy functional of 
Eq.~\eqref{eq:model_MS}, incorrectly predicts a strong relative preference for 
solenoidal supercoils over plectonemes. We believe this could possibly be 
resolved by introducing higher-order terms. In future work, it would be 
interesting to extend the MS model to account for all experimental observables 
even in the high $|\sigma|$ regime. 
Twist-bend coupling could influence, for instance, the long-range transfer
of supercoiling density \cite{vanl12} and the interactions with proteins
which bend and twist DNA \cite{kim93, vand14}.

\begin{acknowledgements}
\paragraph{Acknowledgements} 
We thank M.\ Schurr and F.\ Lanka{\v{s}} for useful discussions and the KU 
Leuven grant IDO/12/08, the Research Funds Flanders (FWO Vlaanderen), the 
Deutsche Forschungsgemeinschaft through Grant No.\ SFB863 for financial 
support.
% the Nanosystems Initiative Munich (NIM) and SFB863 for financial support.
\end{acknowledgements}

% \bibliography{main.bib}


\newpage
\widetext

\section*{Supplemental Material}

%  \title{Supplemental Material for: Twist-bend coupling and the torsional 
%  response of double-stranded DNA}
%  
%  \author{Stefanos K.\ Nomidis}
%  \affiliation{KU Leuven, Institute for Theoretical Physics, Celestijnenlaan 
%  200D, 
%  3001 Leuven, Belgium}
%  
%  \author{Franziska Kriegel}
%  \affiliation{Department of Physics, Nanosystems Initiative Munich, and Center 
%  for NanoScience, LMU Munich, 80799~Munich, Germany}
%  
%  \author{Willem Vanderlinden}
%  \affiliation{KU Leuven, Division of Molecular Imaging and Photonics, 
%  Celestijnenlaan 200F, 3001 Leuven, Belgium}
%  \affiliation{Department of Physics, Nanosystems Initiative Munich,
%  and Center for NanoScience, LMU Munich, 80799~Munich, Germany}
%  
%  \author{Jan Lipfert}
%  \affiliation{Department of Physics, Nanosystems Initiative Munich,
%  and Center for NanoScience, LMU Munich, 80799~Munich, Germany}
%  
%  \author{Enrico Carlon}
%  \affiliation{KU Leuven, Institute for Theoretical Physics, 
%  Celestijnenlaan 200D, 3001 Leuven, Belgium}
%  
%  \date{\today}
%  
% \abbreviations{TWLC, MS, dsDNA, MT, MTT, FOMT, FPA}
% \keywords{DNA, polymers}

%  \maketitle
%  
%  \tableofcontents

\section*{Calculations of $\kappa_b$ and $\kappa_t$ in the MS model}

In this section we present the details of the calculations of $\kappa_b$
and $\kappa_t$, reported in Eqs.(3) amd (4) of the main text.  Defining
$A_1 = A + \varepsilon$ and $A_2 = A - \varepsilon$, we write the energy
of the model as follows
\begin{equation}
\frac{E_\text{MS}}{k_BT} = \frac 1 2 \int_0^L ds \left(
A_1 \Omega_1^2 + A_2 \Omega_2^2 + C   \Omega_3^2 +
2G \Omega_2 \Omega_3 \right)
=
\frac{E_\text{TWLC}}{k_BT} +
\int_0^L ds 
\left[
\frac{\varepsilon}{2} \left( \Omega_1^2 - \Omega_2^2 \right)
+ G \Omega_2 \Omega_3 
\right],
\label{supp:model1}
\end{equation}
where $E_\text{TWLC}$ indicates the energy of the standard TWLC model and the 
two additional terms are the contributions from the bending anisotropy 
($\varepsilon \neq 0$) and the twist-bend coupling ($G \neq 0$). 

As shown in Fig.~1 of the main manuscript, we introduce an
orthonormal set of vectors $\{\vec{\widehat{e}}_1, \vec{\widehat{e}}_2,
\vec{\widehat{e}}_3\}$ which is associated with every point along the
molecule. Here $\vec{\widehat{e}}_3$ is the tangent to the curve and
$\vec{\widehat{e}}_1$ points from the center of the helix towards
its minor groove. The third vector is obtained from the relation
$\vec{\widehat{e}}_2 = \vec{\widehat{e}}_3 \times \vec{\widehat{e}}_1$. In
a relaxed dsDNA molecule the helical axis is completely straight,
corresponding to $\vec{\widehat{e}}_3(s)$ being constant along the
molecule, where $0 \leq s \leq L$ is the arc length.  The double helix
makes a full turn every $l=2\pi/\omega_0 \approx 3.4$~nm, which means
that $\vec{\widehat{e}}_1(s)$ and $\vec{\widehat{e}}_2$(s) are rotated
by an angle of $\omega_0s$ with respect to $\vec{\widehat{e}}_1(0)$ and
$\vec{\widehat{e}}_2(0)$. Any deformation from this ideal state can be
described by the following differential equation~\cite{mark94}
\begin{equation}
\frac{d\vec{\widehat{e}}_i }{ds} =
\left( \omega_0 \vec{\widehat{e}}_3 + \vec{\Omega}  
\right) \times \vec{\widehat{e}}_i, 
\label{eq:de}
\end{equation}
where $|\omega_0 \vec{\widehat{e}}_3 +\vec{\Omega}| \, ds$ is the infinitesimal 
angle around the direction of the vector $\omega_0 \vec{\widehat{e}}_3 + 
\vec{\Omega}$, by which the set $\{ \vec{\widehat{e}}_1, \vec{\widehat{e}}_2, 
\vec{\widehat{e}}_3 \}$ is rotated when going from $s$ to $s+ds$. In general 
$\vec{\Omega}(s)$ depends on the position $s$ and it is customary to introduce 
the three local components as follows $\vec{\Omega} = \Omega_1 
\vec{\widehat{e}}_1 + \Omega_2 \vec{\widehat{e}}_2 + \Omega_3 
\vec{\widehat{e}}_3$.

We now need to express $\{\Omega_i\}$ as functions of the vectors 
$\{\vec{\widehat{e}}_i\}$ and their derivatives. For this purpose we use the 
relations
\begin{equation}
\frac{d\vec{\widehat{e}}_1 }{ds} =
\left( \omega_0 + \Omega_3  \right) \vec{\widehat{e}}_2 
- \Omega_2 \vec{\widehat{e}}_3, 
\label{supp:de1}
\end{equation}
\begin{equation}
\frac{d\vec{\widehat{e}}_2 }{ds} =
\Omega_1 \vec{\widehat{e}}_3 
- \left( \omega_0 + \Omega_3  \right) \vec{\widehat{e}}_1,
\label{supp:de2}
\end{equation}
\begin{equation}
\frac{d\vec{\widehat{e}}_3 }{ds} =
\Omega_2 \vec{\widehat{e}}_1 - 
\Omega_1 \vec{\widehat{e}}_2,
\label{supp:de3}
\end{equation}
which are obtained from Eq.~\eqref{eq:de}. Next we discretize the model 
introducing a discretization length~$a$ and using the following approximations
\begin{align}
\frac{d \vec{\widehat{e}}_i(s)}{ds} &\approx
\frac{\vec{\widehat{e}}_i(s+a)-\vec{\widehat{e}}_i(s)}{a},\\
\vec{\widehat{e}}_i(s) &\approx
\frac{\vec{\widehat{e}}_i(s+a)+\vec{\widehat{e}}_i(s)}{2}.
\end{align}
In order to parametrize the rotation of the frame 
$\{\vec{\widehat{e}}_1(s),\vec{\widehat{e}}_2(s), \vec{\widehat{e}}_3(s)\}$ into 
$\{\vec{\widehat{e}}_1(s+a),\vec{\widehat{e}}_2(s+a), 
\vec{\widehat{e}}_3(s+a)\}$, we introduce three Euler angles $\alpha(s)$, 
$\beta(s)$ and $\gamma(s)$. These angles correspond to a sequence of three 
elementary rotations: one about $\vec{\widehat{e}}_3$, followed by one about 
$\vec{\widehat{e}}_1$ and finally a rotation about $\vec{\widehat{e}}_3$, 
respectively
\begin{equation}
\vec{\widehat{e}}_i (s+a) = \sum_{j=1}^3R_{ij} (\alpha,\beta,\gamma) 
\vec{\widehat{e}}_j (s),
\label{supp:def_rotation}
\end{equation}
where $R$ is the product of three rotation matrices
\begin{equation}\label{supp:Rcomb}
	R = E_3 (\gamma) E_1(\beta) E_3(\alpha)
\end{equation}
with
\begin{equation}\label{supp:Rdef}
	E_1(\phi) = \left(
\begin{array}{ccc}
	1&0&0\\
	0& \cos\phi&\sin\phi\\
	0&-\sin\phi&\cos\phi\\
\end{array}
\right)
\hspace{1cm}\text{and}\hspace{1cm}
E_3(\phi) = \left(
\begin{array}{ccc}
	 \cos \phi&\sin\phi&0\\
	 -\sin\phi&\cos\phi&0\\
	0&0&1\\
\end{array}
\right).
\end{equation}
Plugging this into Eq.~\eqref{supp:Rcomb} gives
\begin{equation}\label{supp:rotEuler}
R = \left(
\begin{array}{ccc}
\cos\alpha \cos\gamma - \sin\alpha \cos\beta \sin\gamma& 
\sin\alpha \cos\gamma + \cos\alpha \cos\beta \sin\gamma& 
\sin\beta \sin\gamma \\
-\cos\alpha \sin\gamma - \sin\alpha \cos\beta \cos\gamma& 
-\sin\alpha \sin\gamma + \cos\alpha \cos\beta \cos\gamma&
\sin\beta \cos\gamma \\
\sin\alpha \sin\beta & -\cos\alpha \sin\beta & \cos \beta\\
\end{array}
\right).
\end{equation}
We can now combine the above equations in order to obtain
\begin{align}
\Omega_1^2 &= \vec{\widehat{e}}_1 \cdot
\frac{d\vec{\widehat{e}}_2 }{ds} \times
\frac{d\vec{\widehat{e}}_3 }{ds} 
= \frac{\vec{\widehat{e}}_1(s+a)+\vec{\widehat{e}}_1(s)}{2}
\cdot
\frac{\vec{\widehat{e}}_2(s+a)-\vec{\widehat{e}}_2(s)}{a}
\times
\frac{\vec{\widehat{e}}_3(s+a)-\vec{\widehat{e}}_3(s)}{a}
\nonumber\\
&= 
\frac{1 + \vec{\widehat{e}}_1(s+a) \cdot \vec{\widehat{e}}_1(s)
- \vec{\widehat{e}}_2(s+a) \cdot \vec{\widehat{e}}_2(s)
- \vec{\widehat{e}}_3(s+a) \cdot \vec{\widehat{e}}_3(s)}{a^2}
= 
\frac{1 + R_\text{11} - R_\text{22} - R_\text{33}}{a^2}
\nonumber\\
&= \frac{\left( 1 - \cos \beta \right)
\left[ 1 + \cos \left(\alpha-\gamma \right)\right]}{a^2},
\\
\Omega_2^2 &= \vec{\widehat{e}}_2 \cdot
\frac{d\vec{\widehat{e}}_3 }{ds} \times
\frac{d\vec{\widehat{e}}_1 }{ds} 
= 
\frac{\left( 1 - \cos \beta \right)
\left[1 - \cos \left(\alpha-\gamma \right)\right]}{a^2}
\label{eq_omega1}
\end{align}
and hence 
\begin{align}
\Omega_1^2 + \Omega_2^2 &= \frac{2}{a^2} \left( 1 -\cos \beta \right),
\label{supp:term1}\\
\Omega_1^2 - \Omega_2^2 &=  \frac{2}{a^2}\left( 1 -\cos \beta \right)
\cos \left( \alpha - \gamma \right).
\label{supp:term2}
\end{align}
The other two terms appearing in Eq.~\eqref{supp:model1} are
\begin{align}
\Omega_3^2 &= \vec{\widehat{e}}_3 \cdot
\left(
\frac{d\vec{\widehat{e}}_1 }{ds} - \omega_0 \vec{\widehat{e}}_2 
\right)
\times
\left(
\frac{d\vec{\widehat{e}}_2}{ds} + \omega_0 \vec{\widehat{e}}_1 
\right)
= \frac{\left( 1 + \cos \beta \right)
\left[ 1 - \cos \left(\alpha+\gamma \right)
-a \omega_0 \sin \left(\alpha+\gamma \right)\right]
+ a^2 \omega_0^2}{a^2},
\\[4pt]
\Omega_2 \Omega_3 &=- \frac{d\vec{\widehat{e}}_3}{ds} \cdot
\left(
\frac{d\vec{\widehat{e}}_2 }{ds} 
+ \omega_0 \vec{\widehat{e}}_1 
\right)
= \frac{\sin \beta 
	\left[2 (\cos \gamma - \cos \alpha) + a \omega_0
\left(\sin \alpha - \sin \gamma\right) \right]
}{2 a^2}.
\end{align}
%%%%%%%%%%%%%%%%%%%%%%%%%%%%%%%%%%%%%%%%%%%%%%%%%%%%%%%%%

In the continuum limit $a \to 0$, the Euler angles become 
infinitesimally small, i.e.\ $\alpha,\beta,\gamma \to 0$. This 
allows us to approximate
\begin{align}
1+\cos \beta &\approx 2,\\
	\cos \alpha+ \frac{a \omega_0}{2} \sin \alpha &\approx 
	\cos \left(\alpha -  \phi_0\right),\\
	\cos \gamma+ \frac{a \omega_0}{2} \sin \gamma &\approx 
	\cos \left(\gamma -  \phi_0\right),\\
\cos \left(\alpha+\gamma \right)
+a \omega_0 \sin \left(\alpha+\gamma \right)
&\approx \cos \left(\alpha+\gamma -2 \phi_0\right),
\end{align}
where we have defined
\begin{equation}
\phi_0 \equiv \frac{a \omega_0}{2} \approx
\sin \left(\frac{a \omega_0}{2} \right)
\end{equation}
and made use of $\cos \phi_0 \approx 1$. With the above approximations we get
\begin{align}
\Omega_3^2 &= \frac{2}{a^2} 
\left[ 1 - \cos \left( \alpha+\gamma-2 \phi_0\right)\right] + \omega_0^2,
\label{supp:term3}
\\
\Omega_2 \Omega_3 &= - \frac{1}{a^2} \sin \beta \left[ 
\cos \left( \alpha - \phi_0 \right)
-
\cos \left( \gamma - \phi_0 \right)
\right].
\label{supp:term4}
\end{align}
Substituting Eqs.~\eqref{supp:term1}, \eqref{supp:term2}, \eqref{supp:term3}
and \eqref{supp:term4} into \eqref{supp:model1} and transforming the 
integral into a sum over segments of length $a$ ($\int_0^L ds \ldots \approx
a \sum_i \ldots $) yields
\begin{align}
\frac{E_\text{MS}}{k_BT} = - \frac{1}{a} \sum_i &\left\{ A \cos \beta_i + C 
\cos
\left( \alpha_i +\gamma_i - 2 \phi_0 \right) -\varepsilon \left( 1- \cos \beta_i 
\right)
\cos \left( \alpha_i -\gamma_i \right) \right.\nonumber\\ 
&\left.\hspace{5pt} +  G \sin \beta_i \left[ \cos
\left( \alpha_i - \phi_0 \right) - \cos \left( \gamma_i - \phi_0 \right)
\right] \right\}, 
\end{align} 
where we have omitted any constant terms. One can simplify this expression by 
introducing the angles $\psi_i\equiv\alpha_i+\gamma_i-2\phi_0$ and 
$\chi_i\equiv\alpha_i-\gamma_i$, 
so as to obtain
\begin{equation}\label{supp:finen}
 \frac{E_\text{MS}}{k_BT}=-\frac{1}{a}\sum_i\left[
 A\cos\beta_i+C\cos\psi_i-\varepsilon(1-\cos\beta_i)\cos\chi_i
 -2G\sin\frac{\chi_i}{2}\sin\frac{\psi_i}{2}\sin\beta_i
 \right],
\end{equation}
where $\beta_i$ and $\psi_i$ are bending and twist angles, respectively.

The total partition function can be written as
\begin{equation}\label{supp:totpart}
 Z=\prod_i\left(\int d\beta_i\sin\beta_id\psi_id\chi_i\right) 
e^{-E_\text{MS}/k_BT} ,
\end{equation} 
where $\beta_i\in[0,\pi]$ and $\psi_i,\chi_i\in[-\pi,\pi]$. 
% From
% Eq.~\eqref{supp:finen} we notice that all segments~$i$ are independent
% from each other, hence it suffices to calculate the partition function
% of a single segment
As the total energy is the sum of independent contributions, it is sufficient 
to consider the partition function of a single segment
%  \begin{align}\label{supp:part0}
\begin{equation}\label{supp:part0}
Z_\text{segm} = \int_0^\pi d\beta \sin\beta \int_{-\pi}^\pi d\psi 
\int_{-\pi}^\pi d\chi \exp \bigg\{\frac{1}{a} \bigg[ 
A\cos\beta+C\cos\psi-\varepsilon(1-\cos\beta)\cos\chi
%  	\nonumber\\
  -2G\sin\frac{\chi}{2}\sin\frac{\psi}{2}\sin\beta
  \bigg]\bigg\}.
%  \end{align}
\end{equation}
We require that the quadratic form \eqref{supp:model1} be positive 
\footnote{A quadratic form in the variable $\Omega_i$ is a homogeneous 
polynomial of degree two in those variables. It can be written in general by  
means of a symmetric matrix $M_{ij}$ as $\sum_{ij} \Omega_i M_{ij} \Omega_j$. 
The quadratic form is said to be positive if the matrix $M$ has positive 
eigenvalues. We require positivity in order for the system to be stable. In this 
case $\Omega_i=0$ corresponds to the minimum value of the form.}, so that the 
minimum of the energy corresponds to a straight ($\beta=0$) and untwisted 
($\psi=0$) conformation. The minimum does not depend on the value of $\chi$. In 
the limit $a\to 0$ we can expand the trigonometric functions in 
Eq.~\eqref{supp:part0} around $\beta=\psi=0$ and extend the integration domains 
of these two variables to~$\infty$
\begin{align}
Z_\text{segm} &\approx e^{(A+C)/a}\int_{-\pi}^{\pi}d\chi
	\int_0^\infty d\beta \beta
 \exp\left[-\frac{\beta^2}{2a}(A+\varepsilon\cos\chi)\right]%\times\nonumber\\
% &\hspace{0.5cm}\times
\int_{-\infty}^\infty d\psi
 \exp\left[-\frac{C}{2a}
 \left(\psi^2+\frac{2G}{C}\sin\frac{\chi}{2}\beta\psi\right)\right]
 \nonumber\\
 &= \ldots
 \frac{e^{(A+C)/a}}{\sqrt{C}}
 \int_{-\pi}^{\pi}d\chi\int_0^\infty d\beta^2
 \exp\left[-\frac{\beta^2}{2a}
 \left(A+\varepsilon\cos\chi-\frac{G^2}{C}\sin^2\!\frac{\chi}{2}
\right)\right]
 \nonumber\\
 &= \ldots 
 \frac{e^{(A+C)/a}}{\sqrt{C}}
\int_{-\pi}^{\pi}d\chi
 \frac{d\chi}{A-G^2/2C+(\varepsilon+G^2/2C)\cos\chi}
%   \nonumber\\
%   &= \ldots 
= \ldots
 \frac{e^{(A+C)/a}}{\sqrt{C}}
 \left[\left(A-\frac{G^2}{2C}\right)^2-\left(\varepsilon+\frac{G^2}{2C}\right)^2 
\right]^{-1/2}\nonumber\\
&= \ldots
 \frac{e^{(A+C)/a}}{\sqrt{C}}
 \left[(A+\varepsilon)\left(A-\varepsilon-\frac{G^2}{C}\right)\right]^{-1/2},
 \label{supp:long_formulas}
\end{align}
where the dots ($\ldots$) denote numerical prefactors which can be ignored, 
since they do not contribute to thermal averages 
\footnote{There is an approximation which was not explicitly mentioned in the 
calculation. In the original partition function $\alpha$ and $\gamma$ vary in 
the domain $-\pi \leq \alpha  \leq \pi$ and $-\pi \leq \gamma \leq \pi$. When 
changing variables to $\psi$ and $\chi$ the integration domain becomes a square 
with the sides tilted of $45^\circ$ with respect to the $\psi$ and $\chi$ axes. 
In Eq.~\eqref{supp:long_formulas} we integrate on $-\infty < \psi < \infty$ and 
$-\pi \leq \chi \leq \pi$. The extension of the integration of $\psi$ to the 
whole real domain is justified by the Gaussian approximation. This is a good 
approximation except at the two ``edges'' of the original integration domain 
$\psi=0$, $\chi=\pm\pi$. In can be shown that the correct calculation produces 
higher-order terms in the discretization length $a$, compared to the result of 
\eqref{supp:long_formulas}.}.

We are interested in the following averages
\begin{equation}\label{supp:cosbeta}
\braket{\cos\beta}=a\frac{\partial}{\partial A}\ln Z_\text{segm} 
=1-\frac{a}{A}\frac{1-\dfrac{G^2}{2AC}}{1-\dfrac{\varepsilon^2}{A^2}-\dfrac{G^2}
{AC}\left(1+\dfrac{\varepsilon}{A}\right)} \equiv 1 - \frac{a}{\kappa_\text{b}}
\end{equation}
and
\begin{equation}\label{supp:coschi}
 \braket{\cos(\alpha+\gamma-2\phi_0)}=a\frac{\partial}{\partial C}\ln 
Z_\text{segm} 
=1-\frac{a}{2C}\frac{1-\dfrac{\varepsilon}{A}}{1-\dfrac{\varepsilon}{A}-\dfrac{
G^2}{AC}} \equiv 1 - \frac{a}{2 \kappa_\text{t}}.
\end{equation}
The difference of a factor $2$ in the definitions above stems from the
differences in the integrations over the Euler angles, with integration
elements $d\beta \sin \beta$ and $d\psi$ for bending and twist,
respectively. We have defined
\begin{equation}\label{supp:lb}
\boxed{
\kappa_\text{b}=
A\frac{1-\dfrac{\varepsilon^2}{A^2}-\dfrac{G^2}{AC}\left(1+\dfrac{\varepsilon}{A
}\right)}{1-\dfrac{G^2}{2AC}}
}
\end{equation}
and
\begin{equation}\label{supp:lt}
\boxed{
\kappa_\text{t}=
C\frac{1-\dfrac{\varepsilon}{A}-\dfrac{G^2}{AC}}{1-\dfrac{\varepsilon
}{A}},
}
\end{equation}
which are Eqs.~(3) and (4) of the main paper. These equations show
that the bending and twist fluctuations between neighboring segments
are governed by \emph{renormalized} bending and torsional stiffnesses
$\kappa_\text{b}$ and $\kappa_\text{t}$. In the TWLC limit $G,\varepsilon
\to 0$ we get $\kappa_\text{b}=A$ and $\kappa_\text{t}=C$, while a
finite twist-bend coupling ($G \neq 0$) gives $\kappa_\text{b}<A$
and $\kappa_\text{t}<C$. This renormalization is induced by thermal
fluctuations, resulting in twisting a thermally fluctuating chain
($\kappa_\text{t}<C$) costing less energy than twisting a straight segment
($\kappa_\text{t}=C$). Note also that a bending anisotropy in absence
of twist-bend coupling ($G=0$ and $\varepsilon \neq 0$) has no effect on
the torsional stiffness ($\kappa_\text{t}=C$), but it modifies bending as
\begin{equation}
\frac{1}{\kappa_\text{b}}=\frac{A}{A^2-\varepsilon^2}
=\frac{1}{2}\left(\frac{1}{A_1}+\frac{1}{A_2}\right),
\end{equation}
i.e.\ the renormalized bending stiffness is the harmonic mean of $A_1$ and
$A_2$~\cite{lank00, esla08}.

%%%%%%%%%%%%%%%%%%%%%%%%%%%%%%%%%%%%%%%%%%%%%%%%%%%%%%%%%%%%%%%%%%%%%%%%%
\begin{figure}[t]
\centering\includegraphics[width=0.82\textwidth]{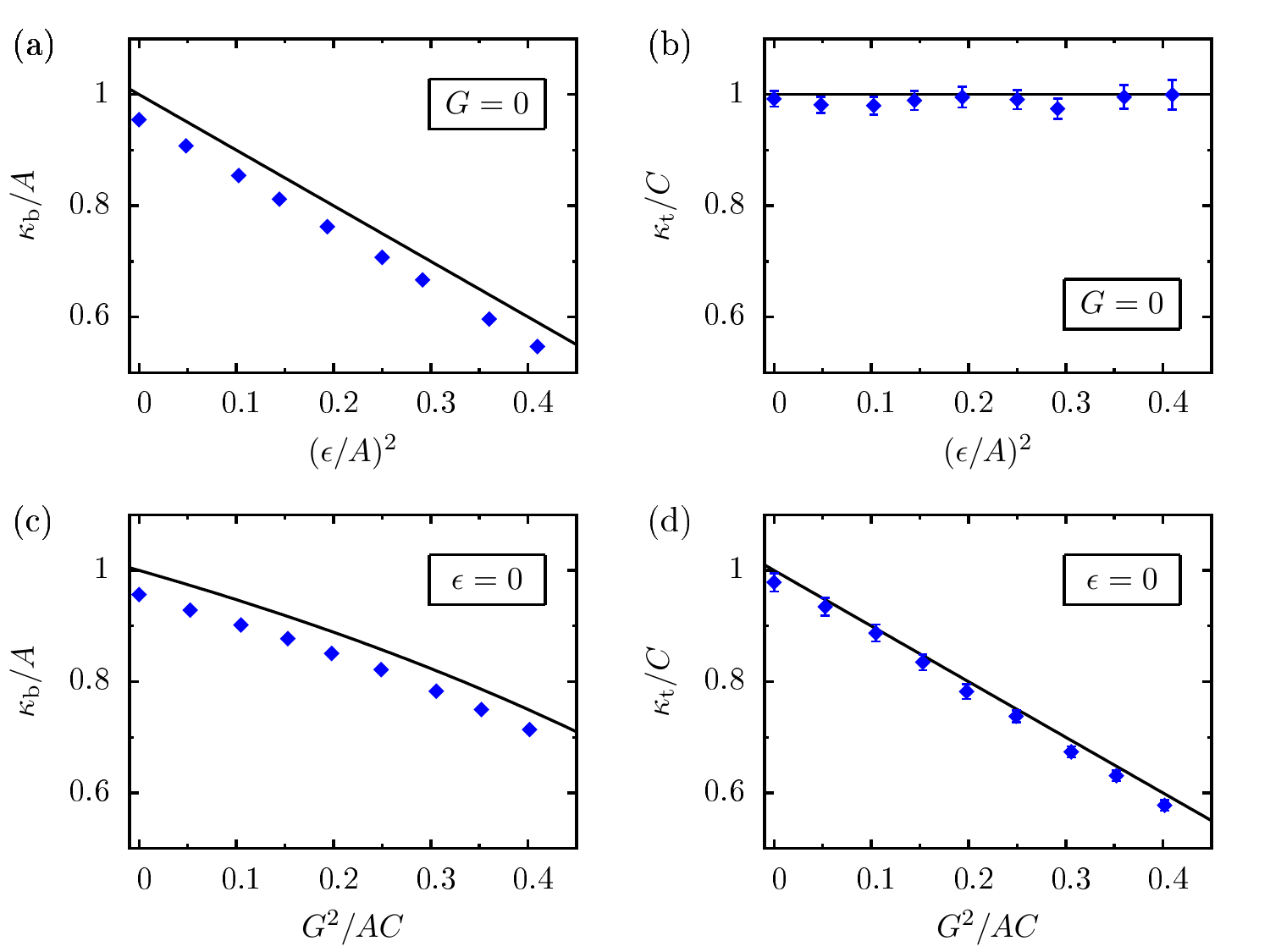}
\caption{Comparison between Eqs.~\eqref{supp:lb} and \eqref{supp:lt}
(black, solid lines) and computer simulations (blue diamonds), showing the
dependence of $\kappa_\text{b}$ and $\kappa_\text{t}$ on the anisotropic
bending (a,b) and twist-bend coupling (c,d).  In both cases theory and
simulations are in good agreement. The error bars are the SEM, and are
not shown where they are smaller than the point size.}
\label{fig:supp1}
\end{figure}
%%%%%%%%%%%%%%%%%%%%%%%%%%%%%%%%%%%%%%%%%%%%%%%%%%%%%%%%%%%%%%%%%%%%%%%%%

Eqs.~\eqref{supp:lb} and \eqref{supp:lt} are exact in the continuum
limit $a \to 0$. Here we compare them with simulations in order to
test the computer model employed for the numerical calculations.  As a
first test, we studied the effect of bending anisotropy and twist-bend
coupling separately. More specifically, we ran simulations for $G=0$
and measured the dependence of $\kappa_\text{b}$ and $\kappa_\text{t}$
on $\varepsilon$. The results are summarized in Figs.~\ref{fig:supp1}(a)
and \ref{fig:supp1}(b), where we also compare with the expressions
\begin{equation}
\frac{\kappa_\text{b}}{A} = 1 - \frac{\varepsilon^2}{A^2}
\hspace{1cm}\text{and}\hspace{1cm}
\frac{\kappa_\text{t}}{C}=1,
\end{equation}
as predicted by Eqs.~\eqref{supp:lb} and \eqref{supp:lt}. We also
tested the dependence on $G$, by setting $\varepsilon=0$ and comparing
with the predictions of Eqs.~\eqref{supp:lb} and \eqref{supp:lt}
\begin{equation}
\frac{\kappa_\text{b}}{A} = \frac{1-\dfrac{G^2}{AC}}{1-\dfrac{G^2}{2AC}}
\hspace{1cm}\text{and}\hspace{1cm}
\frac{\kappa_\text{t}}{C} = 1 - \frac{G^2}{AC}.
\end{equation}
The results are shown in Figs.~\ref{fig:supp1}(c) and
\ref{fig:supp1}(d). In all cases we observe a good agreement between
the two, though the computer model seems to slightly underestimate
$\kappa_\text{b}$ in a systematic way, compared to Eq.~\eqref{supp:lt}. A
possible origin is the continuum-limit approximation that we introduced
in the analytical calculation, as our computer model is discrete.

%%%%%%%%%%%%%%%%%%%%%%%%%%%%%%%%%%%%%%%%%%%%%%%%%%%%%%%%%%%%%%%%%%%%%%%%%
\begin{figure}[t]
\centering\includegraphics[width=0.82\textwidth]{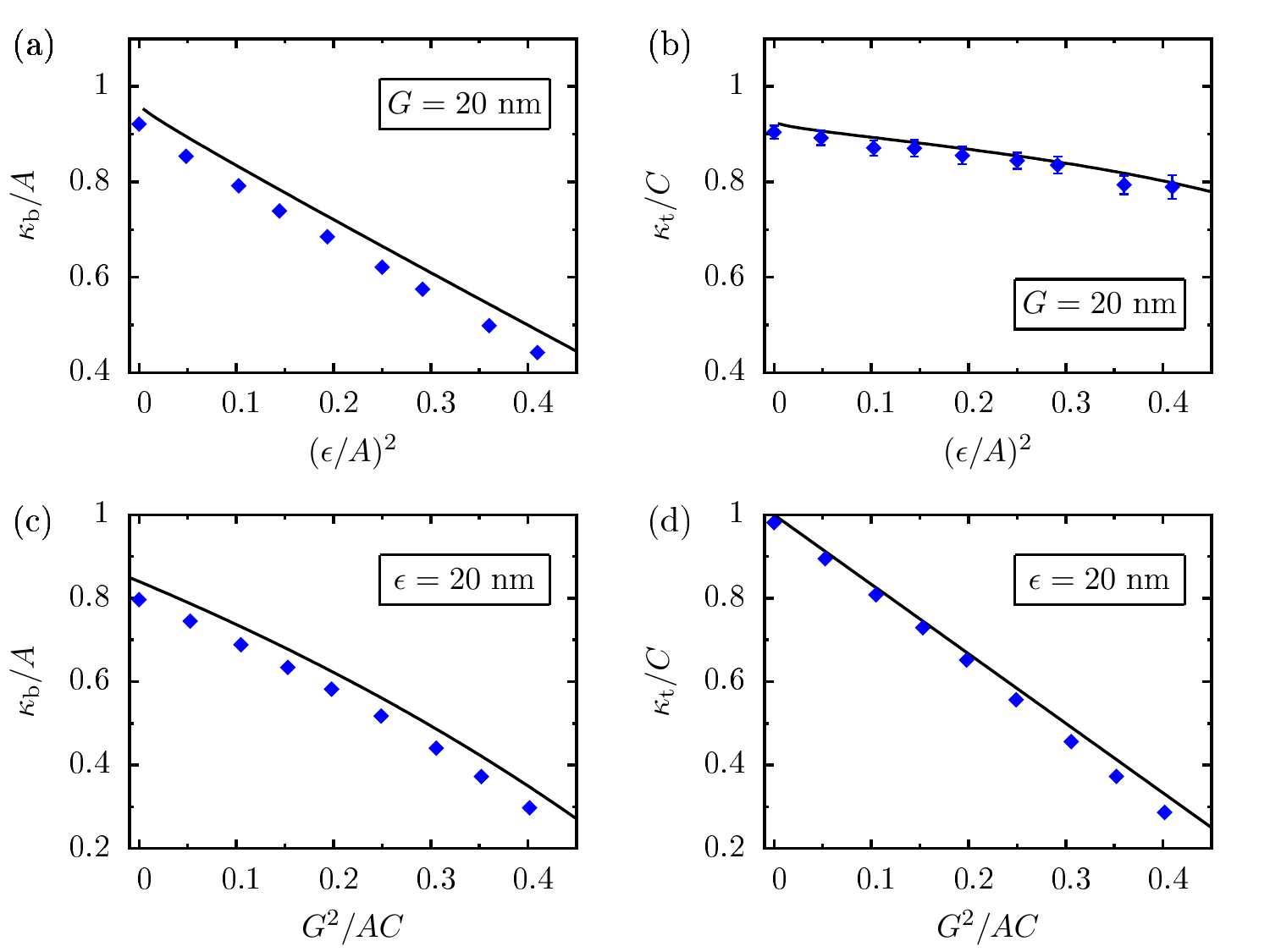}
\caption{Comparison between Eqs.~\eqref{supp:lb} and \eqref{supp:lt}
(black, solid lines) and computer simulations (blue diamonds),
showing how the $\kappa_\text{b}$ and $\kappa_\text{t}$ are affected
when one introduces both the anisotropic bending and the twist-bend
coupling simultaneously. In (a) and (b) we display the dependence of
$\kappa_\text{b}$ and $\kappa_\text{t}$ respectively, on the bending
anisotropy~$\varepsilon$, with the twist-bend coupling constant being
fixed at $G=20$~nm. In a similar manner, in (c) and (d) we fix the bending
anisotropy constant $\varepsilon = 20$~nm and vary $G$. In all cases,
the agreement between theory and simulations is very good. The error bars
are the SEM and is not shown where it is smaller than the point size.}
\label{fig:supp2}
\end{figure}
%%%%%%%%%%%%%%%%%%%%%%%%%%%%%%%%%%%%%%%%%%%%%%%%%%%%%%%%%%%%%%%%%%%%%%%%%

Furthermore, we tested the combined effect of bending anisotropy and
twist-bend coupling, by keeping one of the two properties fixed,
while varying the magnitude of the other. More specifically,
in Figs.~\ref{fig:supp2}(a) and \ref{fig:supp2}(b) we show how
$\kappa_\text{b}$ and $\kappa_\text{t}$ depend on $\varepsilon$,
when setting $G=20$~nm. Similarly, in Figs.~\ref{fig:supp2}(c) and
\ref{fig:supp2}(d) we have taken $\varepsilon=20$~nm and plotted the
$G$-dependence of the persistence lengths. Again, the agreement between
theory and simulations is very good, even under this extreme ``softening''
of the rod.  Thus, we conclude that our computer simulations are in very
good agreement with Eqs.~\eqref{supp:lb} and \eqref{supp:lt}, apart from
a slight systematic deviation in $\kappa_b$.

From Eqs.~\eqref{supp:cosbeta}, \eqref{supp:coschi} one easily obtains the
correlation functions.  For instance, bending correlations are given by
\begin{equation}
\braket{\vec{\widehat{e}}_3(0)\cdot\vec{\widehat{e}}_3(na)}
= \braket{\cos\beta_1 \cos\beta_2 \ldots\cos\beta_n}
= \braket{\cos\beta}^n \equiv e^{-na/l_\text{b}},
\end{equation}
where $l_\text{b}$ is the bending persistence length. 
We then have
\begin{equation}
l_\text{b}\equiv-\frac{a}{\ln\braket{\cos\beta}} = 
-\frac{a}{\ln\left(1 - \frac{a}{\kappa_\text{b}} \right)}
\approx \kappa_\text{b},
\end{equation}
in the limit $a \to 0$.  In a similar manner (see for example
\cite{brac14}) one can define a correlation length associated with
twist as
\begin{equation}
l_\text{t} \approx 2 \kappa_\text{t}
\end{equation}

%%%%%%%%%%%%%%%%%%%%%%%%%%%%%%%%%%%%%%%%%%%%%%%%%%%%%%%%%%%%%%%%%%%%%%%%
\begin{figure}[t]
\centering\includegraphics[width=0.35\textwidth]{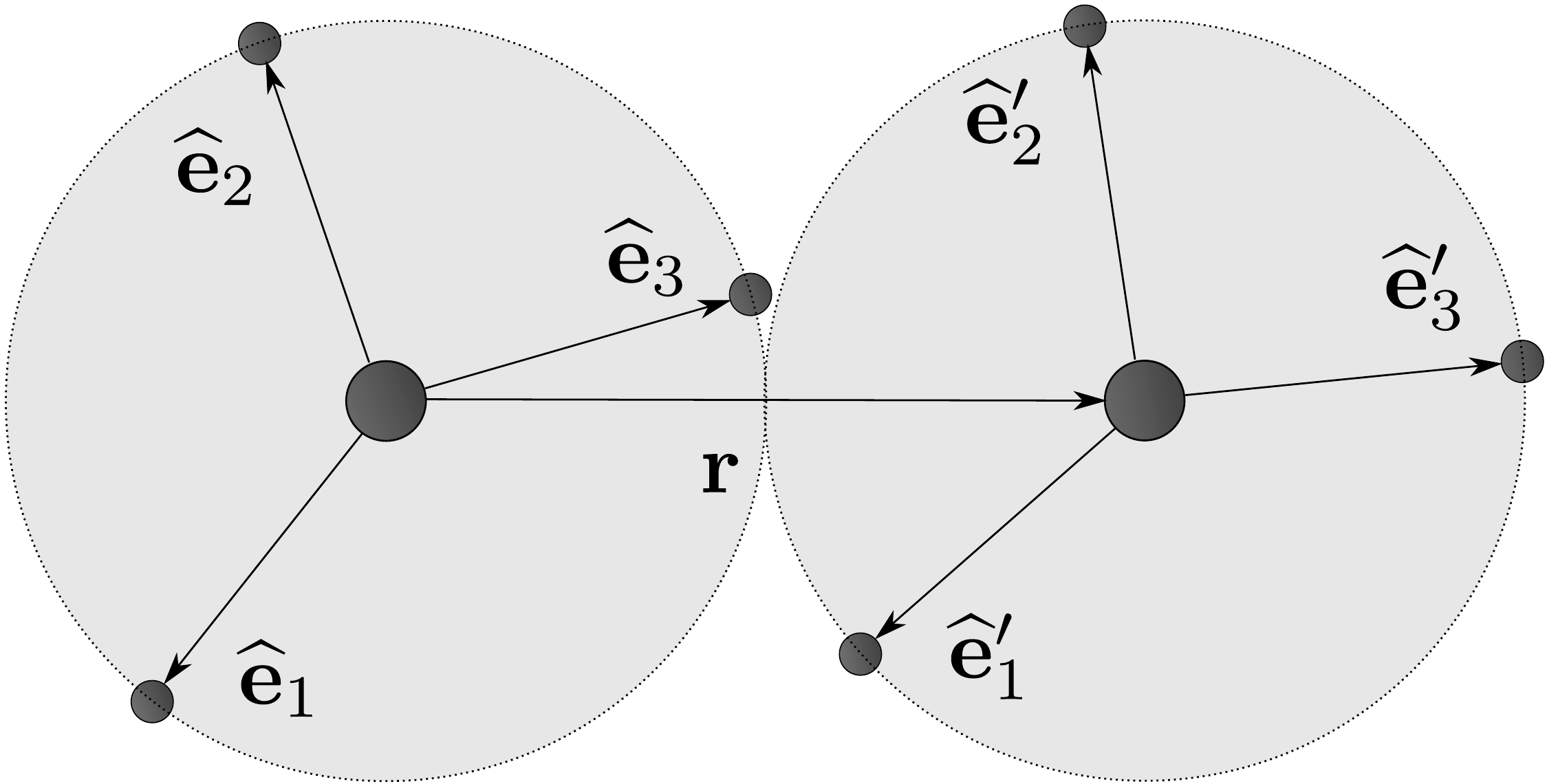}
\caption{An schematic example of the DNA computer model used 
in this work. The molecule is simulated as a series of connected beads, two of 
which are shown. The beads are separated by a distance $|\vec{r}|$, and carry a 
local orthonormal frame $\{ \vec{\widehat{e}}_1, \vec{\widehat{e}}_2, 
\vec{\widehat{e}}_3 \}$, which is represented by three small particles.} 
\label{fig:model}
\end{figure}
%%%%%%%%%%%%%%%%%%%%%%%%%%%%%%%%%%%%%%%%%%%%%%%%%%%%%%%%%%%%%%%%%%%%%%%%

\section*{Coarse-grained computer simulations of DNA}

Computer simulations of the TWLC and MS model were performed using a 
coarse-grained bead-and-spring model, similar to the one discussed in 
Ref.~\cite{brac14}, with the addition of twist-bend coupling and anisotropic 
bending interactions. Each bead is a rigid spherical body, consisting of a core 
particle and three patches at fixed distance from the core defining a local 
frame $\{ \vec{\widehat{e}}_1, \vec{\widehat{e}}_2, \vec{\widehat{e}}_3 \}$. 
Fig.~\ref{fig:model} shows an example of two adjacent beads, together with 
the two respective frames, denoted by $\{ \vec{\widehat{e}}_1, 
\vec{\widehat{e}}_2, \vec{\widehat{e}}_3 \}$ and $\{ \vec{\widehat{e}}_1', 
\vec{\widehat{e}}_2', \vec{\widehat{e}}_3' \}$. These beads are connected via a 
strong finitely extensible nonlinear elastic (FENE) interaction, which keeps 
their separation distance~$r$ very close to a fixed value $|\vec r| \approx a$. 
A very strong interaction term is also used in order to align 
$\vec{\widehat{e}}_3$ with $\vec{r}$ \cite{brac14}, ensuring that 
$\vec{\widehat{e}}_3$ is the local tangent to the polymer chain. The $\Omega_i$ 
terms are computed from a discretization process, as shown in the previous 
section. For instance, the calculation of $\Omega_1^2$ (Eq.~\eqref{eq_omega1}) 
yields
\begin{equation}
\Omega_1^2= 
\frac{1 + \vec{\widehat{e}}'_1 \cdot \vec{\widehat{e}}_1
- \vec{\widehat{e}}'_2 \cdot \vec{\widehat{e}}_2
- \vec{\widehat{e}}'_3 \cdot \vec{\widehat{e}}_3}{a^2}.
\end{equation}
All other terms in the energy functional \eqref{supp:model1} are calculated in a 
similar way, and can be expressed as scalar products between $\{ 
\vec{\widehat{e}}_1, \vec{\widehat{e}}_2, \vec{\widehat{e}}_3 \}$  and $\{ 
\vec{\widehat{e}}'_1, \vec{\widehat{e}}'_2, \vec{\widehat{e}}'_3 \}$. In our 
discretization setup we choose beads with diameter $a=2.3$~nm corresponding to 
$6.7$ base pairs, which is a good compromise between numerical accuracy and 
computational efficiency. There is no intrinsic twist, i.e.\ $\omega_0=0$, in 
the simulations. The two ends of the polymer were attached to an impenetrable 
surface and a large bead, similarly to a typical MT experiment. We included the 
effect of the solvent implicitly, by means of Langevin forces \cite{zwan01}. A 
repulsive Lennard-Jones potential with an effective, hard-core diameter of 
$3.5$~nm \cite{rybe93} was used, in order to avoid distant parts of the polymer 
from overlapping. The effective torsional stiffness was calculated from the 
relation $C_\text{eff} = L / \sigma_\theta^2$, where $L$ is the contour length 
of the polymer ($L=1$~kbp in the simulations) and $\sigma_\theta^2$ is the 
variance of the twist angle. All simulations were performed with the Large-scale 
Atomic/Molecular Massively Parallel Simulator (LAMMPS) \cite{plim95}.

%  \clearpage

%%%%%%%%%%%%%%%%%%%%%%%%%%%%%%%%%%%%%%%%%%%%%%%%%%%%%%%%%%%%%%%%%%%%%%%%%
\begin{figure}[t]
\centering\includegraphics[width=0.45\linewidth]{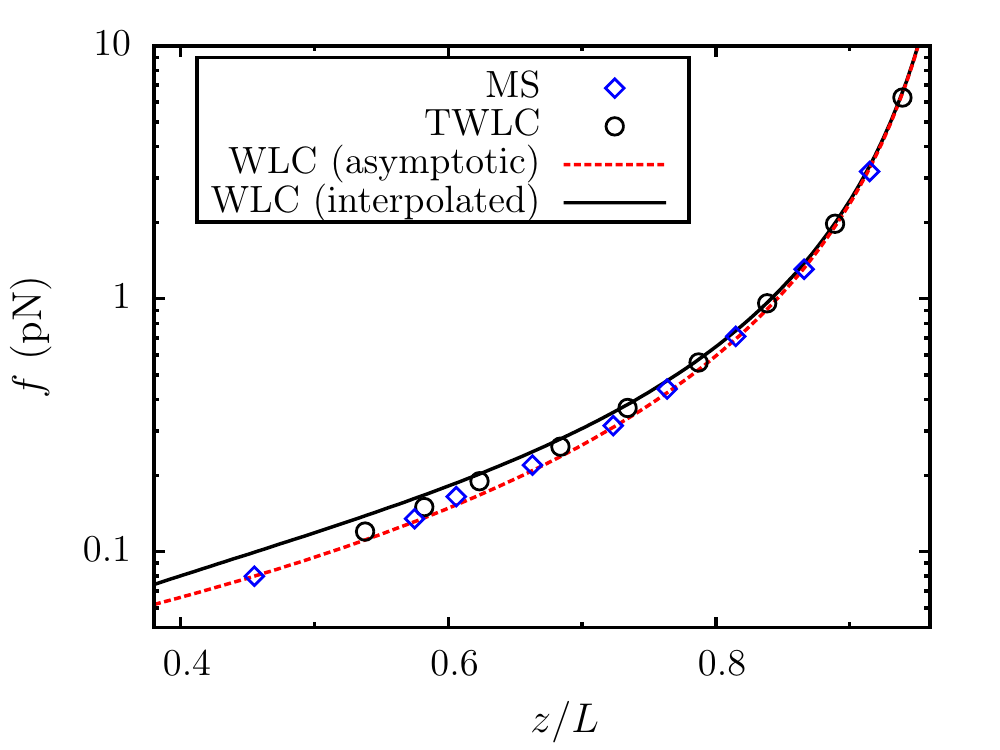}
\caption{Force-extension computer simulations of the TWLC ($A=43$~nm
and $C=110$~nm) and the MS models ($A=56$~nm, $\varepsilon=10$~nm,
$C=110$~nm and $G=40$~nm), together with the interpolated result of
Eq.~\eqref{suppl:fvsext_interpolation} and its high-force limit of
Eq.~\eqref{suppl:fvsext_asymptotic}. Both models are in good agreement
with these expessions. The error bars (SEM) are smaller than the size
of the points and, thus, not shown.}
\label{fig:force_extension}
\end{figure}
%%%%%%%%%%%%%%%%%%%%%%%%%%%%%%%%%%%%%%%%%%%%%%%%%%%%%%%%%%%%%%%%%%%%%%%%%

\section*{Force-extension simulations}

In order to test the used parametrization of the MS model, we performed 
simulations of a dsDNA under an applied, stretching force in the range 
$f=0.08-6.25$~pN and measured its extension in absence of twist 
(Fig.~\ref{fig:force_extension}). We compare between the TWLC with $A=43$~nm 
and 
$C=110$~nm, and the MS model with the values of parameters used throughout this 
work, i.e.\ $A=56$~nm, $\varepsilon=10$~nm, $C=110$~nm and $G=40$~nm. The MS 
model with this parametrization fits the $C_\text{eff}$ data, as shown in the 
main text, and yields a persistence length of $l_b=43$~nm, Eq.~\eqref{supp:lb}. 
Fig.~\ref{fig:force_extension} also plots the asymptotic expression of the WLC 
force-extension curve \cite{mark95}
\begin{equation}
\frac{f l_b}{k_BT} = \frac{1}{4}\, \left(1 - \frac{z}{L}\right)^{-2}
\label{suppl:fvsext_asymptotic}
\end{equation}
and the approximated interpolation formula 
\begin{equation}
\frac{f l_b}{k_BT} = \frac{1}{4}\, \left(1 - \frac{z}{L}\right)^{-2}
- \frac{1}{4} + \frac{z}{L}
\label{suppl:fvsext_interpolation}.
\end{equation}
The latter is known to reproduce within few percents the force-extension 
experimental data in a wide range of forces and is in good agreement with both 
the TWLC and the MS models. The conclusion is that the MS model, with the 
parametrization used throughout this work, is consistent with the measured 
force-extension curves.

%  \clearpage

%%%%%%%%%%%%%%%%%%%%%%%%%%%%%%%%%%%%%%%%%%%%%%%%%%%%%%%%%%%%%%%%%%%%%%%%%
\begin{figure}[b]
\centering\includegraphics[width=0.9\linewidth]{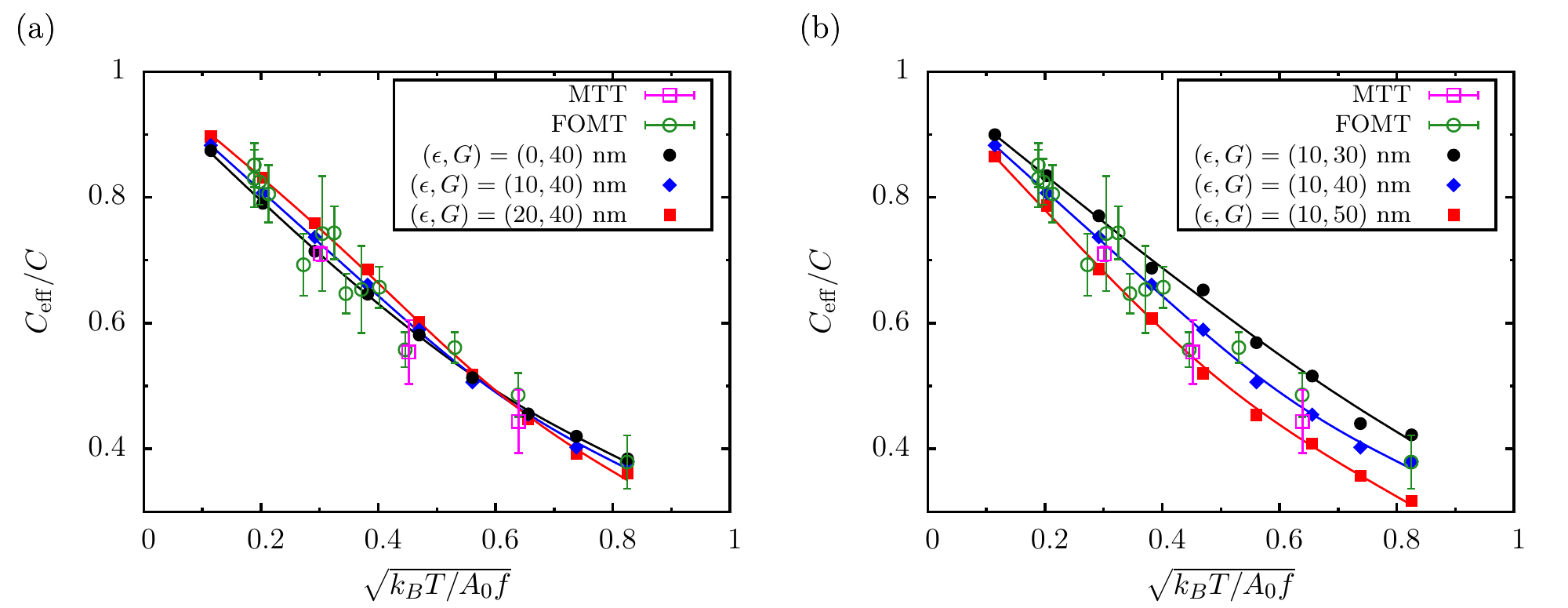}
\caption{Effective torsional stiffness $C_\text{eff}$ obtained from simulations 
of the MS model as a function of the rescaled variable $\sqrt{k_BT/A_0f}$ for 
different values of $G$ and $\varepsilon$. (a) The twist-bend coupling constant 
is fixed to $G=40$~nm, and we vary $\varepsilon = 0$, 10 and 20~nm (reduced 
$\chi^2 = 0.83$, 0.82 and 1.3, respectively). (b) The bending 
anisotropy is fixed to $\varepsilon=10$~nm and we vary $G = 30$, 40 and 50~nm 
(reduced $\chi^2 = 2.1$, 0.82 and 2.5, respectively). The intrinsic 
torsional stiffness is set to $C=110$~nm, while $A$ is fixed by imposing 
$l_b=43$~nm.}
\label{fig:ceff_tests}
\end{figure}
%%%%%%%%%%%%%%%%%%%%%%%%%%%%%%%%%%%%%%%%%%%%%%%%%%%%%%%%%%%%%%%%%%%%%%%%%

\section*{Dependence of $C_\text{eff}$ on $G$ and $\varepsilon$}

We tested the dependence on $G$ and $\varepsilon$ of the effective
torsional stiffness as obtained from simulations of the MS model.
Figure~\ref{fig:ceff_tests} shows the results of the simulations for
(a) fixed $G$ and varying $\varepsilon$ and for (b) fixed $\varepsilon$
and varying $G$. $C_\text{eff}$ depends weakly on $\varepsilon$, while
is much more sensitive to a change of $G$. Figure~\ref{fig:ceff_tests}(b)
shows that the range $30 \leq G \leq 50$~nm fit the experimental
data including error bars, hence our estimate of twist-bend coupling
constant is $G=40\pm 10$~nm.

%  \clearpage

\section*{Dependence of $C_\text{eff}$ on the bending persistence length}

%%%%%%%%%%%%%%%%%%%%%%%%%%%%%%%%%%%%%%%%%%%%%%%%%%%%%%%%%%%%%%%%%%%%%%%%%%
\begin{figure}[t]
\centering\includegraphics[width=0.6\linewidth]{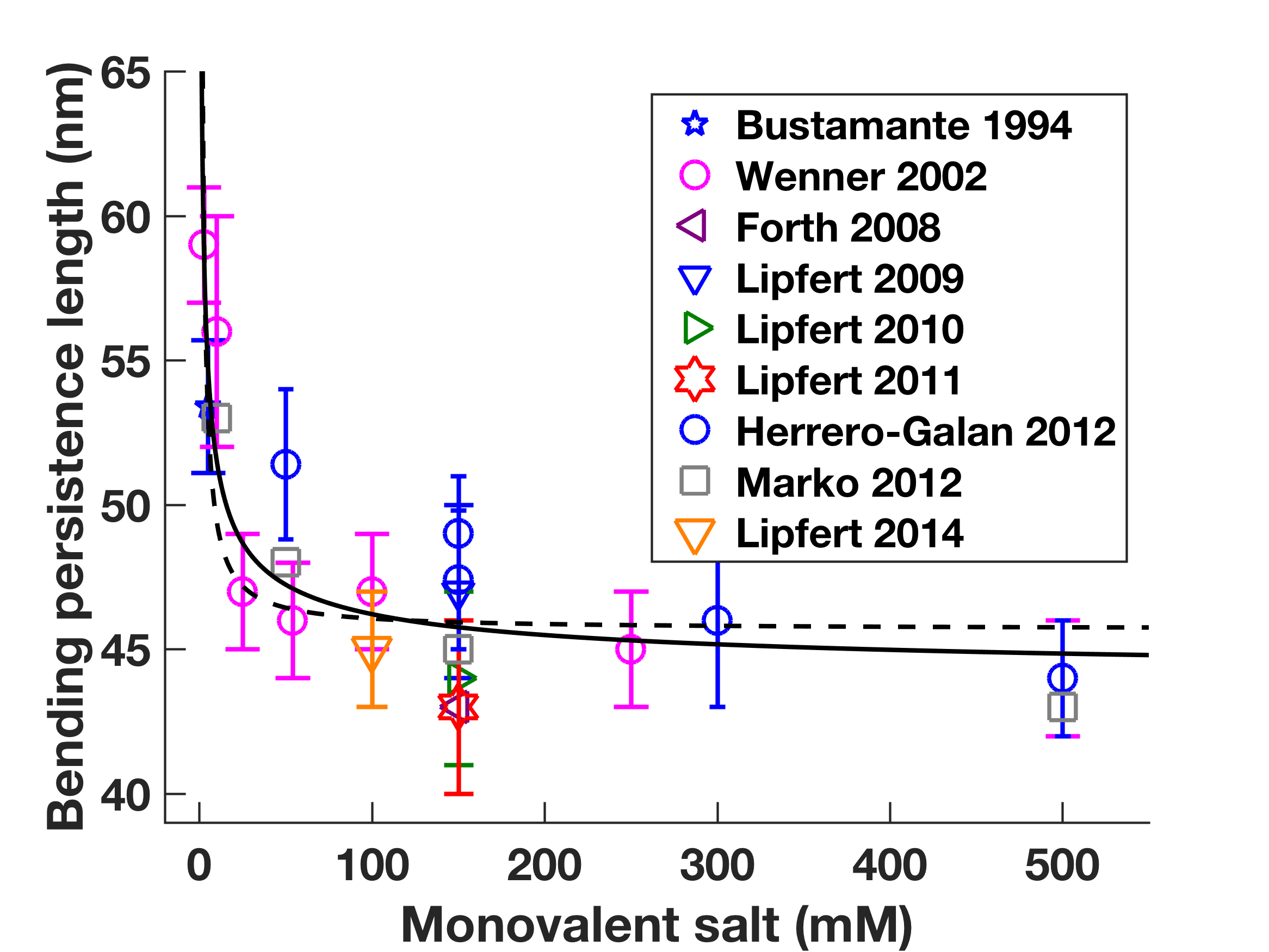}
\caption{Dependence of the bending persistence length of dsDNA on the 
monovalent salt concentration. The plotted data are from Bustamante et 
al.~\cite{bust94}, Wenner et al.~\cite{wenn02}, Forth et al.~\cite{fort08}, 
Lipfert et al.~\cite{lipf09, lipf10, lipf11, lipf14}, Herrero-Galan et 
al.~\cite{herr12} and Marko et al.~\cite{mark15}. The salt dependence can be 
fitted with a function  of the form $\kappa_\text{b} = \kappa_0 + m 
[\text{salt}]^{-\alpha}$,  where $[\text{salt}]$ is the  monovalent salt 
concentration, $\alpha$ is a  scaling  parameter and $\kappa_0$ and  $m$ are 
fitting parameters. According  to the models by Odijk~\cite{odij77} and Skolnick 
and Fixman~\cite{skol77}  it is $\alpha = 1$ (fit shown as dashed line), while 
the model by Barrat  and  Joanny~\cite{barr93} predicts $\alpha = 1/2$ (fit 
shown as solid line). The experiments discussed in the main text are at 
$100$~mM and $150$~mM monovalent salt, which correspond to a bending 
persistence length within the range $43-48$~nm.}
\label{fig:salt}
\end{figure}
%%%%%%%%%%%%%%%%%%%%%%%%%%%%%%%%%%%%%%%%%%%%%%%%%%%%%%%%%%%%%%%%%%%%%%%%%%

The experiments reported in Fig.~2 of the main text are from two independent 
single-molecule approaches: freely orbiting magnetic tweezers (FOMT) 
from Ref.~\cite{lipf11} and magnetic torque tweezers (MTT) from this work. 
The DNA construct was the same in both cases (7.9~kbp, as described in 
Ref.~\cite{lipf10}), but the buffer conditions were slightly different, 
corresponding to 150~mM (FOMT) and 100~mM (MTT) monovalent salt concentration 
(the details of the MTT experiments are discussed below). As shown in 
Fig.~\ref{fig:salt}, the bending persistence length at $100-150$~mM salt lies 
typically within the range $43-48$~nm. For the data shown in the main text the 
bending persistence length was chosen to be $\kappa_\text{b} = 43$~nm (taken 
from Ref.~\cite{lipf11}, obtained from force-extension measurements).

\begin{figure}[t]
\centering\includegraphics[width=0.6\linewidth]{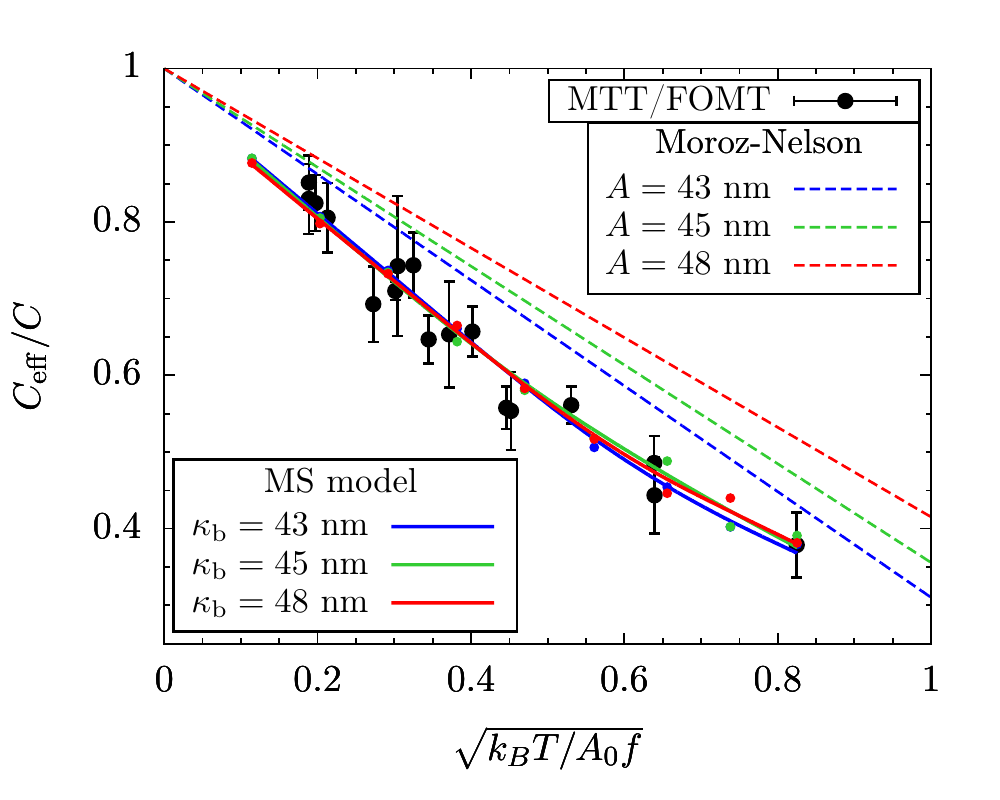}
\caption{Solid lines: simulations of $C_{\rm eff}$ for the MS model using 
different input values of the bending persistence length~$\kappa_\text{b}$.  The 
values we tested (43, 45 and 48~nm with blue, green and red solid interpolated 
lines, respectively) are representative of the experimental conditions 
($100-150$~mM monovalent salt, see Fig.~\ref{fig:salt}) and all yield an 
excellent fit to the MT data (reduced $\chi^2 = 0.74$, $0.66$ and $0.71$, 
respectively). Note that an increase in the persistence length leads to 
stronger deviations between the TWLC prediction (Moroz-Nelson theory, 
Eq.~(5) of main text, plotted with dashed lines) and the experimental data 
(reduced $\chi^2 = 6.1$, $9.0$ and $12.6$ for $\kappa_\text{b} = 43$, $45$ and 
$48$~nm, respectively). As a consequence, the best-fit value of $G$ also 
increases with $\kappa_\text{b}$ ($G = 40$, $43$ and $47$~nm, respectively). In 
all simulations we fixed $C=110$~nm and $\varepsilon=10$~nm (parametrization 
used in the main text), and chose $A$ according to Eq.~\eqref{supp:lb}.}
\label{fig:ceff_lb}
\end{figure}

In Fig.~\ref{fig:ceff_lb} we plot with solid lines the results of
simulations of the MS model, in which the persistence length was fixed
at $\kappa_\text{b} = 43$~nm (as in the main text), $\kappa_\text{b}
= 45$~nm and $\kappa_\text{b} = 48$~nm, while keeping $\varepsilon
= 10$~nm and $C = 110$~nm. Solid lines are the best fit of the MS
model to the experimental data for the given $\kappa_\text{b}$. All
three sets fit equally well the experiments and, as $\kappa_\text{b}$
increases, also the fitted value of $G$ increases (we find $G=43$~nm
and $G=47$~nm for $\kappa_\text{b}=45$~nm and $\kappa_\text{b}=48$~nm,
respectively). Note that an increase in the persistence length leads
to a stronger deviation of the Moroz-Nelson theory, plotted with dashed
lines in Fig.~\ref{fig:ceff_lb}, from the experimental data. Therefore, in
order to fit experiments, one needs a higher correction from twist-bend
coupling (higher $G$) for higher $\kappa_\text{b}$. In conclusion, for
the range of values of $\kappa_\text{b}$ corresponding to the experimental
conditions, the TWLC does not fit the MT data and one needs a relatively
large value of the twist-bend coupling coefficient $G$ to reconcile
theory and experiments.

%%%%%%%%%%%%%%%%%%%%%%%%%%%%%%%%%%%%%%%%%%%%%%%%%%%%%%%%%%%%%%%%%%%%%%%
%%%%%%%%%%%%%%%%%%%%%%%%%%%%%%%%%%%%%%%%%%%%%%%%%%%%%%%%%%%%%%%%%%%%%%%

\section*{Magnetic torque tweezers measurements}

Measurements were performed employing a home-built MT setup and a 7.9-kbp DNA 
construct, as described previously \cite{lipf10}. Specific and torsionally 
constrained coupling of the dsDNA to magnetic beads (1.0~$\upmu$m diameter, 
streptavidin-coated MyOne beads; Life Technologies, USA) and the flow cell 
surface was achieved through ligation of $\sim 600$~bp PCR-generated DNA 
fragments, comprising multiple biotin- and digoxigenin- modified dUTP moieties 
(Jena Bioscience, Germany), respectively, to the central, unmodified DNA. The 
labeled dsDNA molecules were attached to the streptavidin coated beads by 
incubating 5~ng of the DNA construct with 2~$\upmu$L of MyOne beads in a final 
volume of 100~$\upmu$l of phosphate buffered saline (PBS; Sigma-Aldrich, USA) 
for 12~min.

Flow cells were constructed by assembly of two glass coverslips ($24 
\times 60$~mm, Carl Roth, Germany) separated by a single parafilm layer. The 
bottom coverslip was first modified using (3-Glycidoxypropyl)trimethoxysilane 
(abcr GmbH, Germany), subsequently reacted for one hour with anti-digoxygenin 
(100 $\upmu$g/ml in 1$\times$ PBS; Roche, Switzerland) and then passivated 
using BlockAid$^\text{TM}$ Blocking Solution (Thermoscientific) for one hour. 
After flushing of the flow cell with PBS buffer, the DNA-bead solution was 
introduced and allowed to bind for 12~min. Unbound beads are removed from the 
flow cell by flushing with 800~$\upmu$L of PBS buffer. To verify that selected 
beads are bound to a single, torsionally constraint dsDNA tether, several tests 
were performed using a pair of cubic permanent magnets ($5 \times 5 \times 
5$~mm$^3$; W-05-N50-G, Supermagnete, Switzerland), oriented in a horizontal 
configuration above the flow cell. First, the external magnets are moved 
vertically to exert alternating nominal forces of 5~pN and 0.1~pN in order to 
approximately determine the contour length of the tether. Next, magnets are 
rotated counterclockwise by 20 turns at high (5~pN) and low (0.5~pN) applied 
force to identify beads attached via single and fully torsionally constrained 
dsDNA molecules, using the known rotation-extension behavior of dsDNA 
\cite{stri96}. Finally, the flow cell was flushed with $\sim 500$~$\upmu$L of 
TE buffer (10~mM Tris-HCl, 1~mM EDTA, $\text{pH} = 8.0$) supplemented with 
100~mM NaCl, in which the measurements were performed. The relationship between 
magnet height and applied stretching force was determined from the in-plane 
fluctuations by spectral analysis as described previously \cite{te10, lipf14}.

We performed single-molecule torque measurements using our implementation of 
magnetic torque tweezers (MTT), a variant of MT that uses a cylindrical magnet 
with a small additional side magnet to exert a slightly tilted, vertical 
magnetic field \cite{lipf10}. This field configuration provides a weak 
rotational trap for the bead, while applying an upward pulling force. At magnet 
heights of 3, 4 and 5~mm, corresponding to applied forces of 0.9, 0.4 and 
0.2~pN, respectively, we probed the extension and torque response \cite{lipf10, 
lipf14} of the DNA molecules upon changing the linking number in steps of two 
turns, for a total number of 24 turns symmetrically around zero turns, 
corresponding to the torsionally relaxed molecule. Multiple single-molecule 
torque and extension measurements were averaged; the data shown correspond to 
21, 81, and 32 independent molecules for the 0.2, 0.4 and 0.9~pN data, 
respectively.

For the overlay, a shift offset was applied to the extension vs.\ turns
traces, such that the extension-rotation curves are centered around zero
turns for small forces ($<1$~pN). The same shift was applied to the
corresponding molecular torque data. Similarly, a constant extension
offset was applied to the extension data to correct for slightly
different attachment geometries of the DNA to the magnetic beads. As
a consequence, the absolute extension has a larger uncertainty than
the relative extension measurements, which rely on the look-up table
based Z-tracking in the magnetic tweezers with a tracking accuracy of
$\sim1$~nm~\cite{goss02, vilf09}.

%  \clearpage

\section*{Additional extension and torque experiments and simulations}

\begin{figure}[t]
\centering\includegraphics[width=0.7\linewidth]{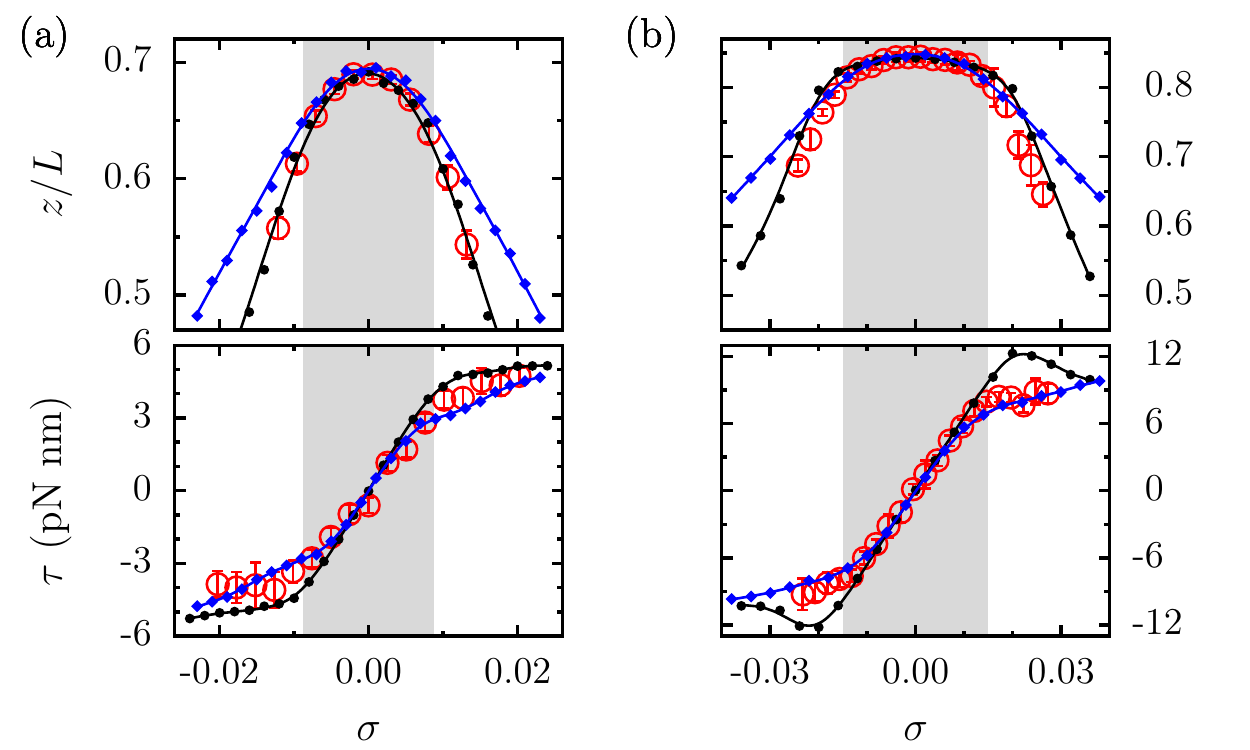}
\caption{Relative extension~$z/L$ and torque~$\tau$, as functions of the 
supercoiling density~$\sigma$, for two different forces (a) $f = 0.2$~pN and 
(b) $f = 0.9$~pN. Open, red circles are from MT experiments, full, black 
circles from TWLC and full, blue triangles from MS simulations. The shaded 
areas correspond to the estimated pre-buckling regime. The figures show
the same features as Fig.~3 of the main text.}
\label{Fig:exttor}
\end{figure}

Besides the extension and torque data presented in Fig.~3 of the main
text, we repeated the experiments and simulations for two different
forces. Figure~\ref{Fig:exttor} shows the additional plots for the two
forces (a) $f=0.2$~pN and (b) $f=0.9$~pN. These data show a similar
behavior to Fig.~3 of the main text and confirm that the torque is
more accurately reproduced by the MS model, whereas the post-buckling
extension agrees with the the TWLC simulation data.

%%%%%%%%%%%%%%%%%%%%%%%%%%%%%%%%%%%%%%%%%%%%%%%%%%%%%%%%%%%%%%%%
%%%%%%%%%%%%%%%%%%%%%%%%%%%%%%%%%%%%%%%%%%%%%%%%%%%%%%%%%%%%%%%%

\clearpage

\section*{Experimental determination of the intrinsic torsional stiffness
$C$ of DNA}

%%%%%%%%%%%%%%%%%%%%%%%%%%%%%%%%%%%%%%%%%%%%%%%%%%%%%%%%%%%%%%%%%%%%%%%%%
\begin{figure}[t]
\centering\includegraphics[width=0.4\textwidth]{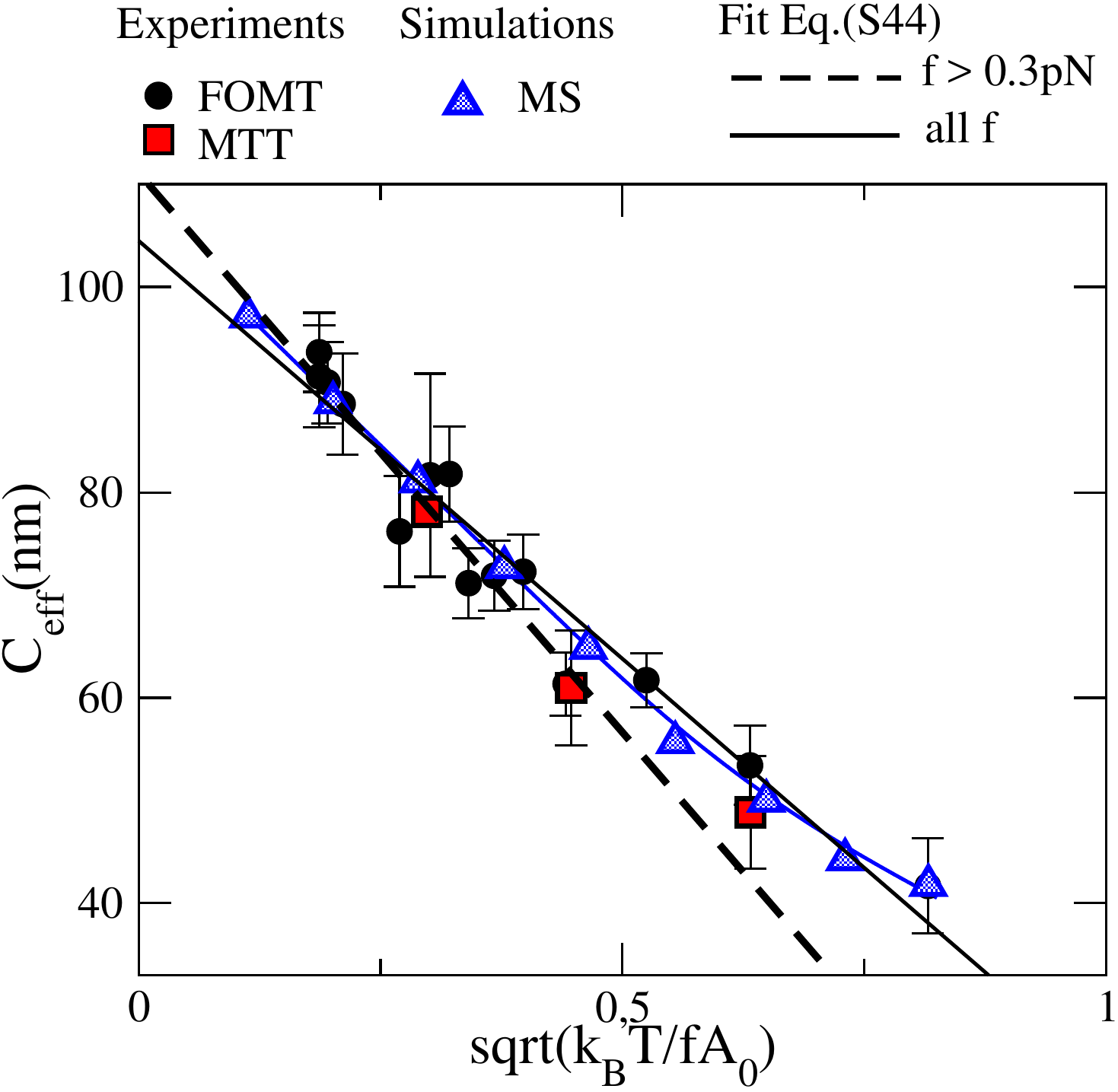}
\caption{Determination of the intrinsic torsional stiffness~$C$
from linear extrapolation of the experimental $C_\text{eff}$ vs. force
data. Fitting a function of the form $C_\text{eff} = C + \Gamma/\sqrt{f}$
to the experimental data (same as in Fig.~(2) of the main text),
allows us to extract $C$. Extrapolations using all data give $C =
105$~nm (solid line), while if we restrict to forces $f>0.3$~pN we get
$C = 110$~nm (dashed line).  Our final estimate is $C=110\pm5$~nm.}
\label{fig:extrapolation}
\end{figure}
%%%%%%%%%%%%%%%%%%%%%%%%%%%%%%%%%%%%%%%%%%%%%%%%%%%%%%%%%%%%%%%%%%%%%%%%%

%%%%%%%%%%%%%%%%%%%%%%%%%%%%%%%%%%%%%%%%%%%%%%%%%%%%%%%%%%%%%%%%%%%%%%%%%%%%%%%
\begin{figure}[ht]
\centering
\includegraphics[width=0.9\linewidth]{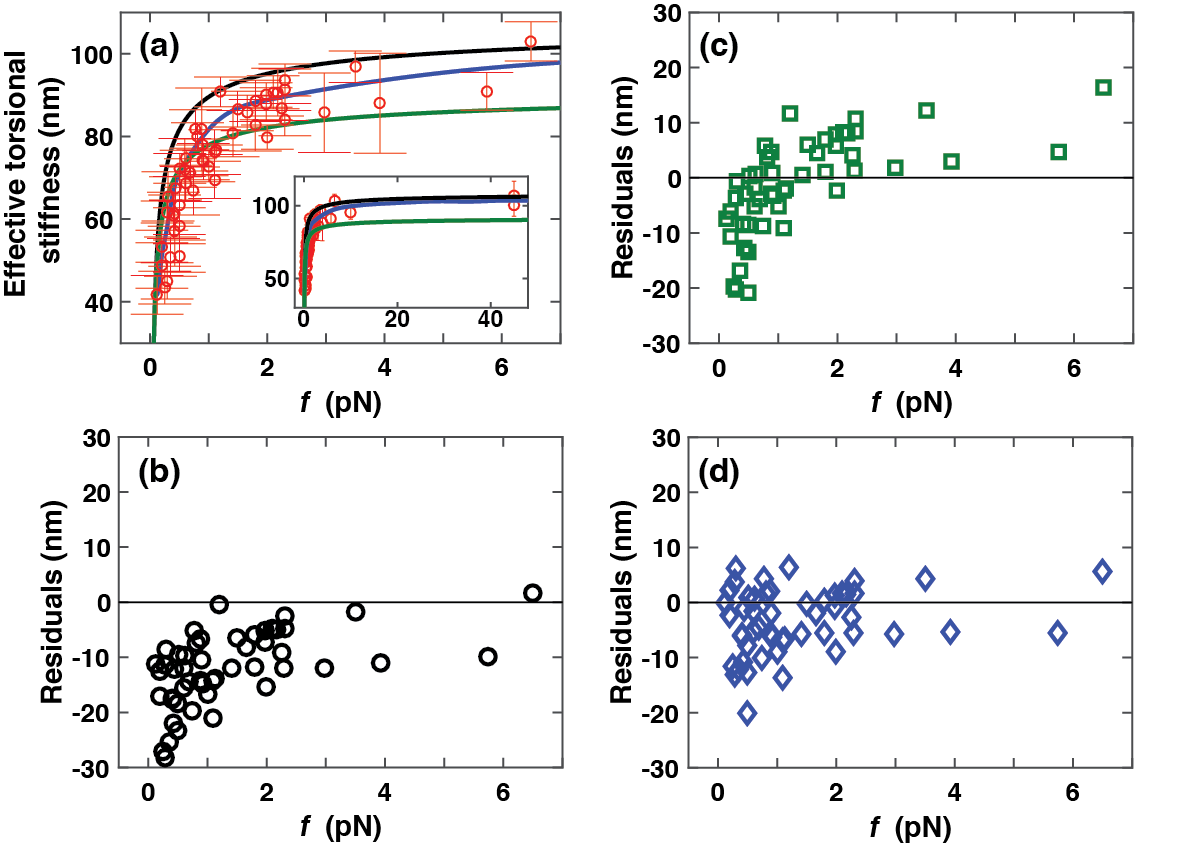}
\caption{Comparison of the TWLC (using the Moroz and Nelson formula,
Eq.~(5) of main text) and the MS model to all available single-molecule
torsional stiffness data~\cite{brya03, oros06, fort08, mosc09, lipf10,
lipf11, kaue11, ober12, lipf14}. (a) Red symbols are all available
single-molecule data on the torsional stiffness of DNA, for which a clear
stretching force can be assigned. The black line is the prediction of
the Moroz and Nelson formula, taken to third order (see~\cite{gore05}),
with $A = 45$~nm  and a (limiting value) $C = 110$~nm (reduced $\chi^2 =
14.4$). The green line is a fit of the Moroz and Nelson formula to the
data with $C$ treated as a fitting parameter, yielding $C = 92.2$~nm
(reduced $\chi^2 = 4.0$). The blue line is the prediction of the MS
model with $G = 40$~nm (see main text; reduced $\chi^2 = 2.9$). (b-d)
Residuals of the fits in panel a, defined as $C_{\text{eff, experiment}}
- C_{\text{eff, theory}}$. (b) Residuals for the Moroz and Nelson formula
with fixed $A$  and $C$. (c) Residuals for the Moroz and Nelson formula
with the (high force) value for $C$ fit to the data. It is apparent that
the residuals do not scatter symmetrically about zero for neither of
the two version of the Moroz and Nelson formula. (d) Residuals for the
MS model with $G = 40$~nm. Not only does this model achieve the lowest
$\chi^2$, but the residuals also scatter clearly more symmetrically
about zero.
} 
\label{fig:c_tests}
\end{figure}
%%%%%%%%%%%%%%%%%%%%%%%%%%%%%%%%%%%%%%%%%%%%%%%%%%%%%%%%%%%%%%%%%%%%%%%%%%%%%%%

Several experimental techniques, such as magnetic and optical tweezers,
cyclization rates and topoisomer distributions have been used in the past
in order to determine the value of the intrinsic torsional stiffness~$C$.
Table~\ref{table} gives a concise overview, with references, of the
measured values from several such studies. As shown in the Table there
is a wide variation in the estimates.

The experimental techniques can be divided into two distinct families,
depending on whether there is a stretching force applied to DNA or not.
One central result of the current study is that the MS model predicts
two distinct values of the torsional stiffness: at high stretching
forces twist is governed by the intrinsic stiffness which we estimate as
$C=110$~nm, while bending fluctuations in an unstretched DNA renormalize
the stiffness to a value $\kappa_\text{t}=75$~nm. This conclusion is
supported by the experimental data reported in Table~\ref{table}.

If the DNA is elongated by a sufficiently strong force, as in magnetic or
optical tweezers, bending fluctuations are suppressed ($\Omega_1 \approx
\Omega_2 \approx 0$) and both the TWLC and the MS models converge to
the twistable rigid rod limit
\begin{equation}\label{eq:TWLC-MS}
\beta E_\text{TWLC} \approx \beta E_\text{MS} \approx \frac{C}{2}\int_0^L 
\Omega_3^2 \, ds - \beta f L,
\end{equation}
where twist stiffness is governed by the parameter $C$.  In practice
one can estimate $C$ from the high-force limit of $C_\text{eff}$.
Figure~\ref{fig:extrapolation} shows an extrapolation based on a
two-parameter fit
\begin{equation}
C_\text{eff} (f) = C + \frac{\Gamma}{\sqrt{f}},
\label{supp:cfit}
\end{equation}
with $C$ and $\Gamma$ being the fitting parameters. As shown in the
figure, it is convenient to plot $C_\text{eff} (f)$ vs. $1/\sqrt{f}$ where
the fit has the shape of a straight line. The analysis yields $C=110\pm5$~nm,
which is the value used throughout the paper.

Extrapolations of $C$ from experimental data sometimes use
the Moroz-Nelson curve
\begin{equation}
C_\text{eff} (f) = C \left( 1 - \frac{C}{4A}\sqrt{\frac{k_BT}{fA}} \right),
\label{supp:MN}
\end{equation}
with $C$ as a free fitting parameter, $A$ being the fixed persistence length. 
This is the procedure used to obtain $C$ from magnetic and optical tweezers data 
reported in the first four rows of Table~\ref{table}. In this fit one assumes 
that the dsDNA is described by the TWLC,  while Eq.~\eqref{supp:cfit} is less 
constraining, assuming only that the asymptotic corrections to $C_\text{eff} 
(f)$ at high forces are of the order $1/\sqrt{f}$.

Fig.~\ref{fig:c_tests}(a) shows a comparison between all available
$C_\text{eff}$ data and the Moroz-Nelson theory of the TWLC, using
the two different fitting procedures (i.e.\ Eqs.~\eqref{supp:cfit}
and \eqref{supp:MN}). We compare them to the predictions of the MS
model, presented in the main text. Calculating the reduced $\chi^2$
value, in combination with plotting the corresponding residuals
(Fig.~\ref{fig:ceff_tests}(b-d)), leads us to the conclusion that the
TWLC cannot account for the experimental data. It is only when using the
MS model, with the high-force extrapolated value of $C$ described above,
that we obtain a quantitative fit to the experimental data.

\clearpage

%%%%%%%%%%%%%%%%%%%%%%%%%%%%%%%%%%%%%%%%%%%%%%%%%%%%%%%%%%%%%%%%%%%%%%%%%
\begin{table}[t]
\centering
\begin{tabular}{c c c c}

 \toprule

 Method & $C$~(nm) & $\kappa_\text{t}$~(nm) \\

 \midrule\\[-0pt]

 $C_\text{eff}$ fit Eq.\eqref{supp:MN} from OT~\cite{fort08} & 100 & & \\[5pt]
 $C_\text{eff}$ fit Eq.\eqref{supp:MN} from OT~\cite{oros06} & 102 & & \\[5pt]
 $C_\text{eff}$ fit Eq.\eqref{supp:MN} from MT~\cite{lipf10} & 109 & & \\[5pt]
 $C_\text{eff}$ fit Eq.\eqref{supp:MN} from MT~\cite{kaue11} & 97 & & \\[10pt]
 
 Extension-rotation curves~\cite{moro97} & 120 & & \\[5pt]
 Extension-rotation curves~\cite{moro98} & 109 & & \\[5pt]
 Extension-rotation curves~\cite{bouc98} & 85 & & \\[10pt]
 
 Stretching under- and overwound DNA~\cite{stri99} & 86 & & \\[5pt]
 High-force $C_\text{eff}$ from RBA~\cite{brya03} & 100-105 & & \\[5pt]
 High-force $C_\text{eff}$ from RBA~\cite{ober12} & 96 & & \\[10pt]
 
 Force-extension of twisted DNA~\cite{volo97} & & 75 & \\[10pt]
 
 Cyclization rates~\cite{shor83a} & & 58 & \\[5pt]
 Cyclization rates~\cite{leve86} & & 83 & \\[5pt]
 Cyclization rates~\cite{tayl90} & & 49 & \\[10pt]
 
 Topoisomer distribution~\cite{shor83b} & & 71 & \\[10pt]
 
 Supercoils free energies~\cite{volo79} & & 74 & \\[10pt]

 FPA~\cite{selv92} & & 46 & \\[5pt]
 FPA~\cite{heat96} & & 53 & \\[10pt]
 
 Spin label~\cite{hurl82} & & 36 & \\[5pt]
 
\bottomrule
\end{tabular}
\caption{Torsional stiffness measured with different techniques.  
Abbreviations used: OT (Optical tweezers), MT (Magnetic tweezers), RBA (Rotor 
bead assay) and FPA (Fluorescence polarization anisotropy).  According to the 
TWLC model all these techniques are expected to measure the intrinsic torsional 
stiffness $C$. According to the MS model, instead, in absence of stretching 
forces and due to bending fluctuations, the torsional stiffness gets 
renormalized to a lower value $\kappa_\text{t} < C$, given by 
Eq.~\eqref{supp:lt}. At strong stretching the bending fluctuations are 
suppressed and the MS model predicts that one should measure $C$. We estimate 
$C=110$~nm and $\kappa_\text{t}=75$~nm. The data in the table are put in two 
different columns, separating experiments sampling twists under stretching 
forces (under the column $C$) and without applied forces (under the column 
$\kappa_\text{t}$). Despite some experimental variability, the data support the 
MS model predictions. Some remarks: Ref.~\cite{volo97} fits force extension 
curves at fixed supercoil density in a region of small tension ($f < 0.5$~pN, 
see Fig.~10) therefore we expect that it samples the renormalized torsional 
stiffness $\kappa_\text{T}$. FPA and spin label techniques estimate the 
torsional stiffness from torsional dynamics, and need as input a model of 
dynamics as well.}
\label{table}
\end{table}
%%%%%%%%%%%%%%%%%%%%%%%%%%%%%%%%%%%%%%%%%%%%%%%%%%%%%%%%%%%%%%%%%%%%%%%%

%  \bibliography{sup}


\end{document}